\documentclass[longauth, traditabstract]{aa} % for the long lists of affiliations 
\usepackage{graphicx}
\usepackage{txfonts}
\usepackage{natbib}
\begin{document}

\newcommand{\OI}{O\,{\sc i}}
\newcommand{\CII}{C\,{\sc ii}}
\newcommand{\CI}{C\,{\sc i}}
\newcommand{\FeII}{Fe\,{\sc ii}}
\newcommand{\SiII}{Si\,{\sc ii}}
\newcommand{\SI}{S\,{\sc i}}
\newcommand{\NeII}{Ne\,{\sc ii}}

\def\kms{{\hbox {km\thinspace s$^{-1}$}}} 
\def\Lsun{{\hbox {$L_{\odot}$}}} 
\def\Msun{{\hbox {$M_{\odot}$}}} 

\title{The complete far-infrared and submillimeter
spectrum of the Class~0 protostar Serpens SMM1  obtained with
\textit{Herschel}\thanks{{\em Herschel} is an ESA space
observatory with science instruments provided by European-led
Principal Investigator consortia and with important participation
from NASA.}}
\subtitle{Characterizing UV-irradiated shocks heating and chemistry}

\author{Javier R. Goicoechea\inst{1},
J.~Cernicharo\inst{1}, 
A.~Karska\inst{2,3}, 
G.J.~Herczeg\inst{4}, 
E.~T.~Polehampton\inst{5,6},\\
S.~F.~Wampfler\inst{7},
L.~E.~Kristensen\inst{3},
E.F.~van~Dishoeck\inst{2,3},
M.~Etxaluze\inst{1},
O.~Bern\'e\inst{8,9},
R.~Visser\inst{10}}

\institute{Centro de Astrobiolog\'{\i}a (CSIC/INTA), Ctra. de Torrej\'on a Ajalvir, km 4
28850, Torrej\'on de Ardoz, Madrid, Spain. \\
\email{jr.goicoechea@cab.inta-csic.es}
\and
Max-Planck-Institut f\"ur extraterrestrische Physik (MPE), Postfach 1312, 85741 Garching, Germany.
\and
Leiden Observatory, Leiden University, PO Box 9513, 2300 RA Leiden, The Netherlands. 
\and
Kavli Institute for Astronomy and Astrophysics, Peking University, Yi He Yuan Lu 5, HaiDian Qu,  100871 Beijing, China. 
\and
RAL Space, Rutherford Appleton Laboratory, Chilton, Didcot, Oxfordshire, OX11 0QX, UK.
\and
Institute for Space Imaging Science, University of Lethbridge, 4401 University Drive, Lethbridge, Alberta, T1J 1B1, Canada.
\and
Centre for Star and Planet Formation, Natural History Museum of Denmark, 
University of Copenhagen, \O{}ster Voldgade5-7, DK-1350 K\o{}benhavn K, Denmark.
\and
Universit\'e de Toulouse; UPS-OMP; IRAP;  Toulouse, France.
\and
CNRS; IRAP; 9 Av. colonel Roche, BP 44346, F-31028 Toulouse cedex 4, France
\and
Department of Astronomy, University of Michigan, 500 Church St., Ann Arbor, MI 48109, USA
}

\date{Submitted 28 June 2012; accepted 17 September 2012}

\abstract{
We present the first complete $\sim$55-671\,$\mu$m  spectral scan  of a low-mass Class~0 protostar
(Serpens~SMM1) 
taken with the PACS and SPIRE spectrometers on board \textit{Herschel}.
More than 145 lines have been detected, most of them rotationally excited lines of $^{12}$CO 
(full ladder from $J_u$=4-3 to 42-41 and $E_{\rm u}$/$k$$=$4971\,K), 
H$_2$O (up to 8$_{18}$-7$_{07}$ and $E_{\rm u}$/$k$$=$1036\,K), OH (up to $^2\Pi_{1/2}$ $J$=7/2-5/2 and  $E_{\rm u}$/$k$$=$618\,K), 
$^{13}$CO (up to $J_u$=16-15), HCN and HCO$^+$ (up to $J_u$=12-11). Bright [\OI]63, 145\,$\mu$m and weaker 
[\CII]158 and [\CI]370, 609\,$\mu$m lines are also detected, but
excited lines from chemically related species (NH$_3$, CH$^+$, CO$^+$, OH$^+$ or H$_2$O$^+$) are  not.
Mid-IR spectra retrieved from the \textit{Spitzer} archive are also first discussed here.
The $\sim$10-37\,$\mu$m spectrum has many fewer lines, but shows clear detections of [\NeII],
[\FeII], 
[\SiII] and [\SI] fine structure lines,  as well as weaker H$_2$ $S$(1) and $S$(2) pure  rotational lines.  
The observed line luminosity is dominated by CO ($\sim$54\%), H$_2$O ($\sim$22\%), [\OI] ($\sim$12\%)
and OH ($\sim$9\%) emission. 
A  multi-component radiative transfer model allowed us
to approximately quantify the contribution of the three different temperature components suggested by the $^{12}$CO rotational ladder
($T$$_{\rm k}^{hot}$$\approx$800\,K, $T$$_{\rm k}^{warm}$$\approx$375\,K
 and $T_{\rm k}^{cool}$$\approx$150\,K). 
Gas densities  $n$(H$_2$)$\gtrsim$5$\times$10$^{6}$\,cm$^{-3}$  are needed to reproduce
the observed far-IR lines arising from shocks in  the inner protostellar envelope (\textit{warm} and \textit{hot} components)
 for which we derive upper limit  abundances of 
$x$(CO)$\lesssim$10$^{-4}$, $x$(H$_2$O)$\lesssim$0.2$\times$10$^{-5}$ and $x$(OH)$\lesssim$10$^{-6}$ 
 with respect to H$_2$. 
The lower energy submm $^{12}$CO and H$_2$O lines show more extended emission 
that we associate with the \textit{cool}  entrained outflow gas.
Fast dissociative $J$-shocks (v$_{\rm s}>60$\,km\,s$^{-1}$) within an embedded atomic jet,
as well as lower velocity small-scale non-dissociative  shocks (v$_{\rm s}\lesssim20$\,km\,s$^{-1}$)  
are needed  to explain both the atomic 
fine structure lines and  the \textit{hot} CO and H$_2$O lines respectively.
Observations also show the signature of UV radiation (weak [\CII] and [\CI] lines  
and high HCO$^+$/HCN abundance ratios) and thus, most observed species  
likely arise in UV-irradiated shocks. 
Dissociative $J$-shocks produced by a jet impacting the ambient material are the most
probable origin of [\OI] and OH emission and of a significant fraction of the \textit{warm} CO emission.
In addition, H$_2$O photodissociation in UV-irradiated non-dissociative shocks along the outflow cavity walls
can also contribute to the [\OI] and OH emission.

}

\titlerunning{The complete spectrum of the Class 0 protostar Serpens SMM1}
	\authorrunning{Goicoechea et al.}
\maketitle
%
%________________________________________________________________

\section{Introduction}

In the earliest stages of evolution,  low-mass protostars
are deeply embedded in dense envelopes of molecular gas and dust.
During collapse, conservation of angular momentum combined with infall along the magnetic
field lines leads to the formation of a dense rotating protoplanetary disk
that drives the accretion process. At the same time, both mass and
angular momentum are removed from the system by the onset of
jets, collimated flows, and the magnetic braking action \citep[\textit{e.g.,}][]{Bon96, Bac99}.
The resulting  outflow produces a cavity in the natal envelope with walls subjected to energetic shocks and 
strong radiation fields from the protostar. 
These processes heat the circumstellar gas and leave their signature in the prevailing chemistry 
(\textit{e.g.,}~enhanced abundances of water vapour).

Both the line and continuum emission from Young Stellar Objects (YSOs) peak in the far-infrared
(far-IR) and thus they are robust diagnostics of the energetic processes
associated with the first  stages of star formation \citep[\textit{e.g.,}][]{Gia01}.
\begin{figure*}[ht]
\begin{center}
\includegraphics[width=14.5cm, angle=0]{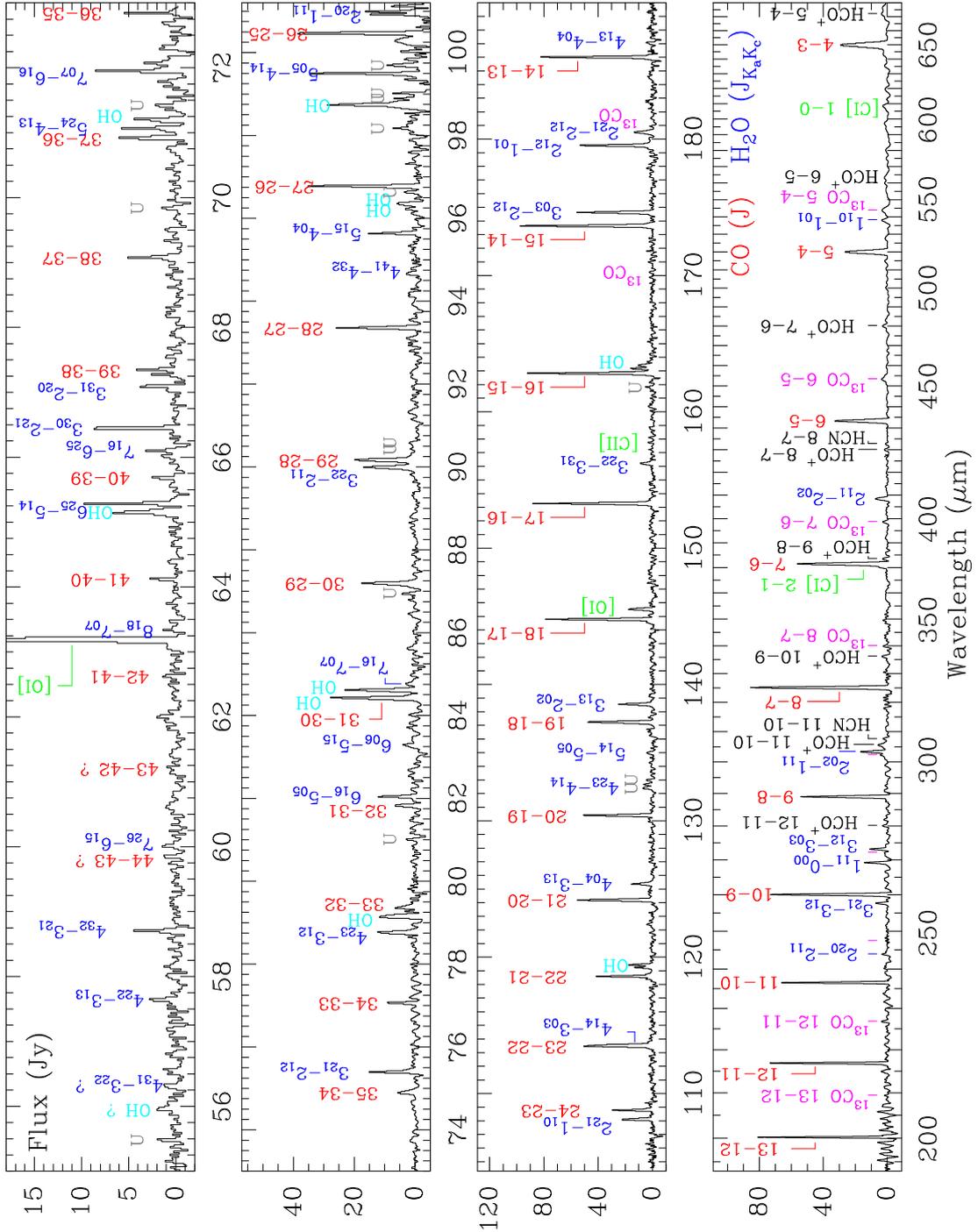}
\caption{PACS and SPIRE (lower panel) continuum-subtracted spectra of 
Serpens SMM1 (central spaxel) with
H$_2$O (blue), CO (red), OH (cyan), $^{13}$CO (magenta),
[\OI], [\CII] and [\CI]  (green), and HCO$^+$ and HCN (black) lines labelled. The flux density scale
is in Jy (assuming a point source).} 
\end{center}
\label{fig_obs_spectra}
\end{figure*}
The improved sensitivity and angular/spectral resolution of \textit{Herschel} spectrometers compared to previous far-IR observatories
allows us to detect a larger number of far-IR  lines and to better constrain their spatial origin. 
This is especially true for the detection of high excitation, optically thin lines that can help us to identify high temperature components as well as to disentangle and quantify the
dominant
heating mechanisms (mechanical vs. radiative). Early \textit{Herschel} observations  of a few low-mass protostars for
which partial or complete PACS spectra are available show that the molecular emission of relatively excited $^{12}$CO 
($J$$<$30), H$_2$O and OH lines is a common feature
(van Kempen et al. 2010 for HH46 and   Benedettini et al. 2012 for L1157-B1).
This \textit{warm} CO and H$_2$O emission was suggested to arise from the walls of an outflow-carved cavity in the
envelope, which are heated by UV photons and by non-dissociative $C$-type shocks 
\citep[][]{vK10,Vis12}.
The [\OI] and OH emission observed in these sources, however, was proposed to arise from dissociative $J$-shocks.
Herczeg et al. (2012) detected higher-$J$  $^{12}$CO lines (up to $J$=49-48) and
highly excited H$_2$O lines in the NGC~1333 IRAS~4B outflow indicating the presence of even hotter gas.
The very weak [\OI]63\,$\mu$m line emission in IRAS~4B  outflow (the [\OI]145\,$\mu$m line is  not even detected)
led these authors to conclude that the hot gas where H$_2$O dominates the gas cooling is heated by 
non-dissociative $C$-shocks shielded from UV radiation.
Passive heating by the protostellar luminosity is also thought to contribute to the mid-$J$ 
$^{12}$CO and $^{13}$CO emission \citep{Vis12,Yil12}. In NGC~1333 IRAS~4A/4B, however, the 
$^{12}$CO intensities and broad line-profiles of lower-$J$ transitions ($J$=1-0 up to 10-9) probe 
swept-up or entrained shocked gas along the outflow  \citep{Yil12}.

\begin{figure}[t]
\begin{center}
\resizebox{\hsize}{!}{\includegraphics[angle=-0]{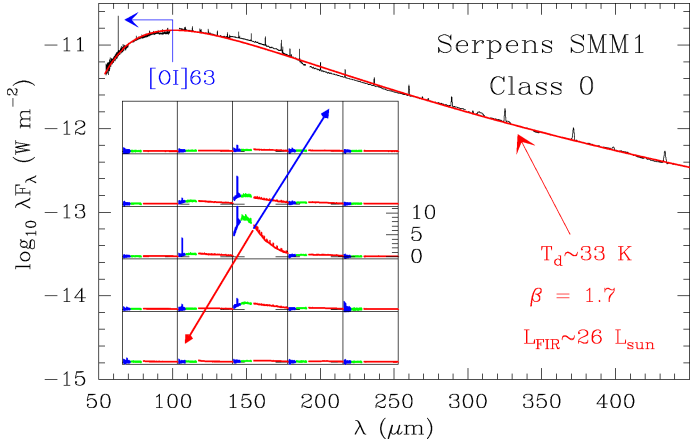}} 
\caption{Observed SED of Serpens SMM1 and 
modified blackbody fit (red curve). 
\textit{Inset} shows the full PACS  array  in flux density units of 
10$^{-14}$\,W\,m$^{-2}$\,$\mu$m$^{-1}$. The abscissa is in linear scale from 55 to 190\,$\mu$m.  
The bright emission line seen in several positions is [\OI]63\,$\mu$m. Approximate outflow  directions are shown
with red and blue arrows.} 
\end{center}
\label{fig_sed}
\end{figure}

In this paper we present the first complete far-IR and submm spectra of a Class~0 protostar 
(Serpens\,SMM1 or FIRS\,1), taken with the PACS \citep{Pog10} and SPIRE \citep{Grif10} spectrometers on board
the \textit{Herschel Space Observatory} \citep{Pil10}. 
SMM1 is a  low-mass protostellar  system
located in the Serpens core \citep{Eir08} at a distance of $d$=230$\pm$20\,pc \citep[see also][for an alternative measurement]{Dzi10}. 
It is the most massive low-mass Class~0 source ($M_{\rm env}$$\simeq$16.1\,M$_{\odot}$, $T_{\rm bol}$$\simeq$39\,K) 
of the WISH program sample
\citep[\textit{Water in Star-forming Regions with Herschel},][]{vD11}.
The \textit{Herschel} spectra are complemented with mid-IR \textit{Spitzer} Infrared Spectrograph (IRS)
archive spectra of the embedded protostar \citep[photometry first presented by][]{Eno09}.

The Serpens cloud core 
was first mapped in the far-IR with the LWS spectrometer on board the \textit{Infrared Space Observatory} 
(ISO). Even with a poor angular and spectral resolution of $\sim$80$''$ and $R$=$\lambda$/$\Delta\lambda$$\sim$200, ISO-LWS observations
of Serpens\,SMM1 revealed
a rich spectrum of molecular lines \citep{Lar02}, superposed onto a strong  far-IR  continuum
\citep{Lar00}. In particular, $^{12}$CO (up to $J$=21-20), H$_2$O (up to 4$_{41}$-3$_{30}$) and some OH 
 lines were detected. Detailed modeling of the ISO emission concluded that those species
trace the inner $\sim$10$^3$\,AU regions around the protostar, where gas temperatures are above 300\,K and densities above
10$^6$\,cm$^{-3}$. The comparison of the H$_2$O, OH and CO lines with
existing shock models suggested that shock heating along the  SMM1 outflow is
the most likely mechanism of molecular excitation. The details of these shocks, however, were less clear.

\begin{figure}[t]
\begin{center}
\resizebox{\hsize}{!}{\includegraphics[angle=-0]{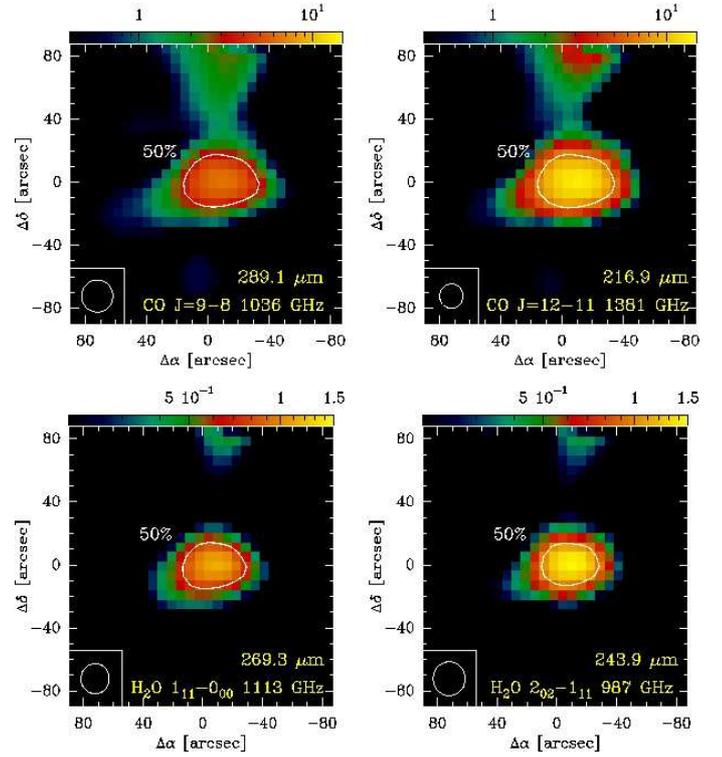}}
\end{center}
\caption{\textit{Upper}: SPIRE-FTS $^{12}$CO $J$=9-8 and 12-11 
sparse maps. 
\textit{Lower}: $p$-H$_2$O 1$_{11}$-0$_{00}$ and 2$_{02}$-1$_{11}$ maps respectively.
The line surface brightness  is on the 10$^{-8}$\,W\,m$^{-2}$\,sr$^{-1}$ scale. The white contours represent
the 50\% line emission peak level.
The FWHM beam  is shown in each inset. 
Note the increase of line surface brightness  in the more excited lines.} 
\label{fig_spiremap}
\end{figure}

Because of its  far-IR  luminosity and rich spectrum, Serpens~SMM1 is an ideal Class~0 source
for studying the protostellar environment
and to characterize the shock- and UV-heating and chemistry with the much higher spatial
and spectral capabilities provided by \textit{Herschel}. 
In the following sections we present the data set, we identify the observed lines, 
present a 3 component non-LTE radiative transfer model  and compare the observations
with available shock models and observations of other YSOs.

\vspace{-0.2cm}
\section{Observations and Data Reduction}

\subsection{Herschel PACS and SPIRE Observations}

PACS spectra between $\sim$55 and $\sim$210\,$\mu$m were obtained on 31 October 2010
in the Range Spectroscopy mode.
The PACS spectrometer uses photoconductor detectors and  provides 25 spectra  
over a 47$''$$\times$47$''$ field-of-view (FoV) resolved in 5$\times$5 spatial pixels (``spaxels''), 
each with a size of $\sim$9.4$''$ on the sky.
The resolving power varies between $R$$\sim$1000 (at $\sim$100\,$\mu$m in the R1
grating order) 
and $\sim$5000 (at $\sim$70\,$\mu$m in the B3A order). The central spaxel was centered at 
the Serpens SMM1 protostar  
($\alpha_{2000}$: 18$^h$29$^m$49.8$^s$, $\delta_{2000}$: 1$^o$15$'$20.5$''$). 
Observations were carried out in the  ``chop-nodded'' mode with the largest chopper throw of 6~arcmin.
The total observing time was $\sim$3\,h (observation IDs 1342207780 and 1342207781). 
The measured width of the spectrometer point spread function (PSF)  is relatively constant for
$\lambda$$\lesssim$100\,$\mu$m (FWHM$\simeq$spaxel size) but increases at longer wavelengths.
In particular only $\simeq$40\%\,($\simeq$70\%)  of a point 
source emission would fall in the central spaxel at $\simeq$190\,$\mu$m ($\lambda$$\simeq$60\,$\mu$m). 
Hence, extracting line 
surface brightness from a semi-extended source like Serpens\,SMM1 is not trivial. 
In this work we followed the method described by Herczeg et al. (2012) and Karska et al. (submitted).
This involves multiplying all line fluxes measured in the central spaxel with a $\lambda$-dependent
 \textit{correction curve}
calculated from the  extended emission of the brightest CO and H$_2$O lines observed 
over the entire FoV.
The correction factor is constant ($\sim$2) for $\lambda$$<$120\,$\mu$m  
and increases from  $\sim$2 to $\sim$3-4 at $\sim$200\,$\mu$m. 
For the continuum and [\OI] and  [\CII]  lines, 
the individual line fluxes measured in all spaxels were  added.
Note that we checked that the reference spectra in the two off positions are free of [\CII] emission to
confirm that the  [\CII] lines  towards Serpens\,SMM1 are real.
Figure~1 shows the PACS spectrum in the central spaxel 
and Fig.~2 shows the entire PACS array from 55 to 190\,$\mu$m 
(the bright emission line is [\OI]63\,$\mu$m).
\begin{figure*}[t]
\begin{center}
\vspace{-0.0cm}
\includegraphics[width=14.8cm]{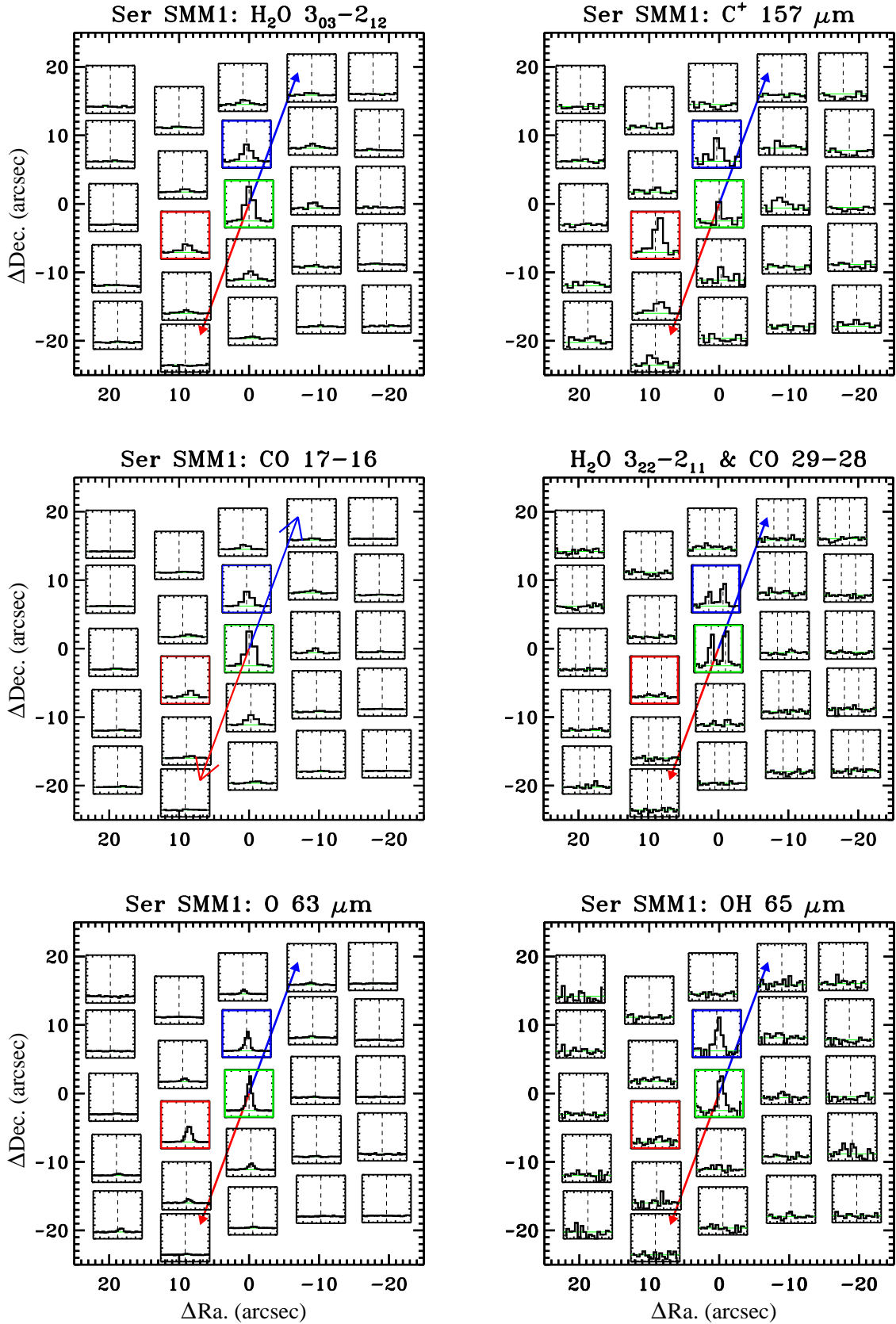}
\caption{PACS spectral maps of Serpens\,SMM1 in the $o$-H$_2$O 3$_{03}$-2$_{12}$ (174.626\,$\mu$m);
[\CII]157.741\,$\mu$m; CO $J$=17-16 (153.267\,$\mu$m); both $p$-H$_2$O 3$_{22}$-2$_{11}$ (89.988\,$\mu$m) and 
CO $J$=29-28 (90.163\,$\mu$m); [\OI]63.183\,$\mu$m and OH $^{2}\Pi_{3/2}$ $J$=9/2$^+$-7/2$^-$ (65.279\,$\mu$m) lines
(from \textit{top} to \textit{bottom}).
The center of each spaxel corresponds to its offset position  with respect to the protostar at
$\alpha_{2000}$: 18$^h$29$^m$49.8$^s$, $\delta_{2000}$: 1$^o$15$'$20.5$''$.
The Y-axis for each transition represents the normalized line flux (in the  -0.2 to 1.2 range)
with respect to the brightest line in the array.  
The X-axis represents the -550 to 550\,km\,s$^{-1}$ velocity scale except for the
OH 65.279\,$\mu$m line (-350 to 350\,km\,s$^{-1}$) and the $p$-H$_2$O 89.988\,$\mu$m and CO 90.163\,$\mu$m lines
(-1000 to 1000\,km\,s$^{-1}$). The vertical dashed lines show the position of the rest wavelengths.
Blue and red arrows show the approximated outflow direction. 
The green box represents the central spaxel centered at the protostar position whereas the red and blue boxes
represent postions in the red and blue outflow lobes respectively.
Note that the PSF 
is relatively constant for $\lambda$$\lesssim$100\,$\mu$m (FWHM$\simeq$spaxel size$\simeq$9.4$''$) 
but increases at longer wavelengths.} 
\end{center}
\label{fig_pacsmaps}
\end{figure*}

\vspace{+0.2cm}
SPIRE FTS spectra between $\sim$194 and $\sim$671\,$\mu$m (1545-447\,GHz) 
were obtained on 27 March 2011. The SPIRE FTS uses two bolometer arrays covering 
the 194-313\,$\mu$m  (Short Wavelength array, SSW) and 303-671\,$\mu$m  (Long Wavelength array, SLW) bands.
The two arrays contain 19 (SLW) and 37 (SSW) hexagonally packed detectors separated by 
$\sim$2 beams (51$''$ and 33$''$ respectively). The unvignetted FoV is $\sim$2$'$.
The  observation was centered at Serpens SMM1 (IDs 1342216893) in the high spectral 
resolution mode ($\Delta\tilde{\lambda}$=0.04\,cm$^{-1}$; $R$$\sim$500-1000). 
\clearpage
\begin{figure}[h]
\begin{center}
\vspace{-0.0cm}
\resizebox{\hsize}{!}{\includegraphics[angle=0]{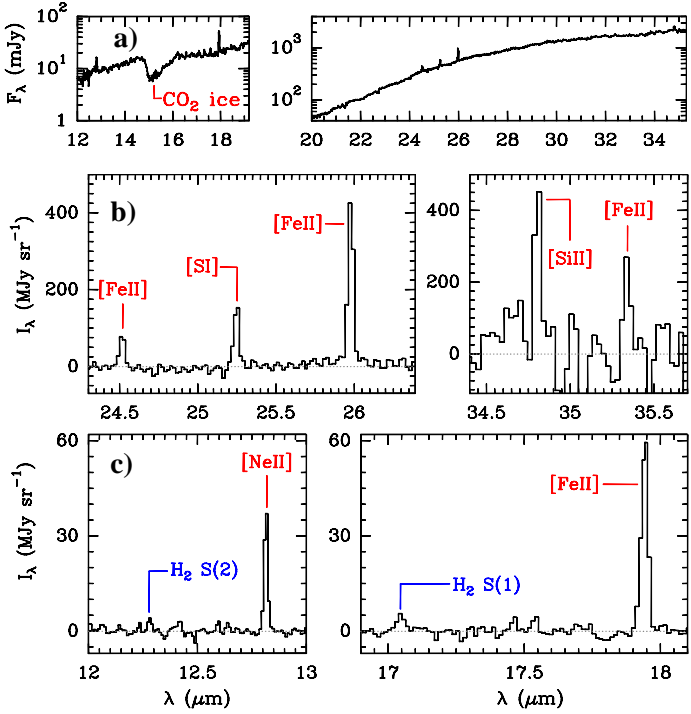}}
\caption{$(a)$ \textit{Spitzer} IRS spectrum of the Class~0 protostar Serpens SMM1  using 
the Hi Res modules and a 7$''$$\times$7$''$ aperture.
$(b)$ Continuum-subtracted spectra with main narrow lines identified in the
LH module ($\lambda\geq$19.5$\mu$m) and $(c)$ with the SH module ($\lambda$$\leq$19.5$\mu$m).
Note the lack of bright molecular line emission (H$_2$O, OH, etc.).} 
\end{center}
\label{fig:irs}
\end{figure}
The total integration time was $\sim$9\,min.
We corrected the spectrum to account for the extended continuum and molecular emission
of the source and the changing beam size with frequency across
the SPIRE band. Taking account of the coupling of the source
to the SPIRE-FTS beam (Wu et al. in prep.) and using the overlap region between the SSW
and SLW bands as a reference (where the beam size differs by a factor
of $\sim$2) we calculate the FWHM size of the submm continuum
source to be $\sim$20$''$. The SPIRE continuum fluxes were appropriately
corrected and the unapodized spectrum was used to fit the line intensities with sinc functions.
Four representative mid-$J$ $^{12}$CO and H$_2$O  integrated line intensity maps are shown in Figure~3.

SPIRE and PACS data were 
processed using HIPE  and then exported to GILDAS where basic line spectrum
manipulations were carried out. Tables A.1 and A.2 summarize the detected lines
and the line fluxes towards the protostar position (central spaxel; corrected for extended emission in the case
of PACS observations). The spatial distribution of several atomic and molecular
lines observed in the PACS array is shown in Figure~4.

\subsection{Spitzer IRS Observations}

In order to complement the far-IR and submm \textit{Herschel} spectra we have also analyzed mid--IR data  towards
the protostar position from the Infrared Spectrograph (IRS) on board \emph{Spitzer}~\citep{Hou04} in 
high spectral resolution mode ($R$$\sim$600; 9.9-37.2\,$\mu$m). 
The Basic Calibrated Data (BCD) files for the Short High (SH) and Long High (LH) orders of IRS were retrieved 
from the \textit{Spitzer} Heritage Archive\footnote{http://sha.ipac.caltech.edu/applications/Spitzer/SHA/} 
(OBSID 34330, PI M. Enoch). Pipelining of the data was achieved using CUBISM \citep{Smi07} to assemble 
the BCD files into spectral cubes. Bad and rogue pixels were removed using 3-sigma filtering. The spectra
were extracted in a $\sim$7$''$$\times$7$''$ aperture, thus close to the PACS spaxel size.
The resulting \textit{Spitzer} IRS spectra are shown in Figure~5.
The photometry of these data was first presented by Enoch et al. (2009).
Table~A.3 summarizes the detected lines
and their fluxes. 

\vspace{-0.2cm}
\section{Results}

\subsection{Far-Infrared and Submillimeter Lines}

Figure~\ref{fig_obs_spectra} shows the very  rich PACS and SPIRE  (lower panel) 
spectrum of the Serpens~SMM1 protostar.
More than 145 lines have been detected, most of them rotationally excited lines from abundant molecules:
38 $^{12}$CO lines (up to $J$=42-41 and $E_{\rm u}$/$k$$=$4971\,K), 37 lines of both 
$o$-H$_2$O and  $p$-H$_2$O (up to 8$_{18}$-7$_{17}$ and $E_{\rm u}$/$k$$=$1036\,K), 
16  OH lines (up to $^2\Pi_{1/2}$ $J$=7/2-5/2 and $E_{\rm u}$/$k$$=$618\,K),
12 $^{13}$CO lines (up to $J$=16-15 and $E_{\rm u}$/$k$$=$719\,K) and several HCN and HCO$^+$ lines 
(up to $J$=12-11 and $E_{\rm u}$/$k$$=$283\,K).
Weaker [\CII]158\,$\mu$m and [\CI]370, 609\,$\mu$m lines 
are also detected. Excited lines from  NH$_3$, CH$^+$, CO$^+$, OH$^+$ or H$_2$O$^+$
are however not detected at the PACS and SPIRE spectral resolutions and sensitivity.
The brightest line in the spectra is the [\OI]63\,$\mu$m line (see also Figure~2)
with  a luminosity of $L_{\rm 63}$$\simeq$0.014\,L$_{\odot}$. Assuming no extinction in the far-IR, we measure 
[\OI]63\,$\mu$m/[\CII]158\,$\mu$m$\simeq$30  and  [\OI]63\,$\mu$m/[\OI]145\,$\mu$m$\simeq$10 line flux ratios
integrated over the entire PACS array.

Most of the far-IR continuum and line  emission arises from the protostar position (central spaxel)
with weaker, but detectable, contributions from adjacent outflow positions (see Figure~4). 
In particular, the high excitation lines of CO, H$_2$O and OH detected below $\lambda$$<$100\,$\mu$m seem to show a more
compact distribution than the lower excitation lines (within the PACS PSF sampling caveats).
The highest PACS spectral resolution  ($\sim$90-60\,km\,s$^{-1}$) is achieved in the $\sim$60-70\,$\mu$m
range. Although at this resolution all detected molecular lines  are spectrally unresolved (they have narrower intrinsic 
line-widths), the observed [\OI]63\,$\mu$m line is $>$30\,\% broader than any close-by molecular line in the $\sim$60-70\,$\mu$m
range. This likely indicates that PACS  marginally resolves the [\OI]63\,$\mu$m line, being broader (or having
higher velocity line-wing emission) 
than the excited far-IR CO, H$_2$O and OH lines.  
In addition, the [\OI]63\,$\mu$m line-profile peak shifts in velocity from the protostar position (where
the line is brightest)  to the
outflow red-lobe position  where the line peak is  redshifted by  $\sim$100\,km\,s$^{-1}$ 
(more than a resolution element) and the molecular line emission is weaker (see red boxes in Figure~4).
 Note that [\CII]158\,$\mu$m 
peaks at this red-lobe position and the line peak is also redshifted by $\sim$100\,km\,s$^{-1}$ 
(although less than a resolution element in this wavelength range).
Similar [\OI]63\,$\mu$m velocity-shifts are also seen in sources such as HH46 and are associated
with the emission from an atomic jet itself \citep{vK10}. 
We refer to Karska et al. (submitted) for a detailed
comparison of the [\OI]63\,$\mu$m shifts in several outflows.

Finally, the lower energy submm lines, from $^{12}$CO and H$_2$O  in particular, show a more extended 
distribution than the far-IR lines, with some indication of outflow emission in the the S-E direction 
(see SPIRE maps in Figure~3).

Table~1 summarizes the total line luminosities adding all observed lines between 
$\sim$55 and 671\,$\mu$m.  
Note that a few unidentified lines (U) are present in the PACS spectrum.

\begin{figure}[ht]
\begin{center}
\vspace{-0.4cm}
\resizebox{\hsize}{!}{\includegraphics[angle=-90]{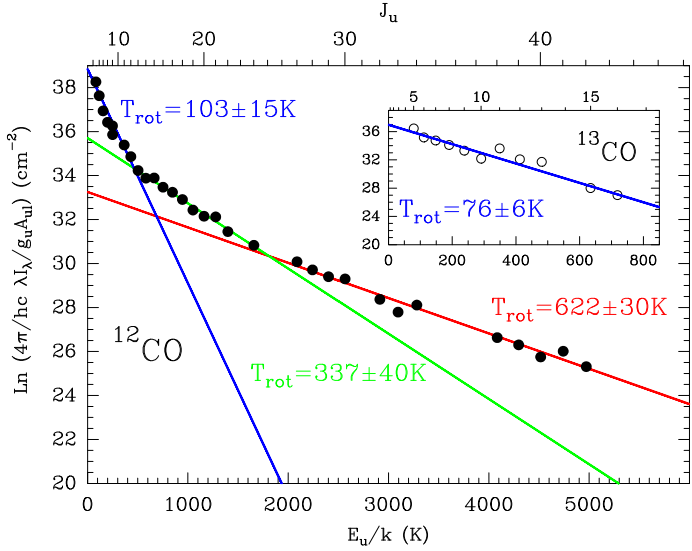}}
\caption{$^{12}$CO and $^{13}$CO rotational diagrams obtained from PACS and SPIRE spectra. 
The estimated error in $T_{\rm rot}$ for $^{12}$CO corresponds to different choices of $J_u$ cutoffs 
for the different components. The range of possible $T_{\rm rot}$ for $^{13}$CO reflects the standard
error on the fitted values.} 
\end{center}
\label{fig_rotdia}
\end{figure}

\subsection{Mid-Infrared Lines and Extinction}

The  mid-IR spectrum of Serpens SMM1 (Figure~5$a$) shows a much less rich spectrum than in the far-IR 
domain (probably due to severe dust extinction and lack of spectral resolution). 
Nevertheless, two weak H$_2$ pure rotational lines (0-0~$S$(1) and $S$(2) transitions) and seven brighter atomic
 fine-structure lines of Ne$^+$, Si$^+$, S and Fe$^+$ (four transitions)
are clearly detected\footnote{First ionization potential of observed
ionic species are: 21.56 (Ne), 11.26 (C), 10.36 (S), 8.15 (Si) and 7.90\,eV (Fe).}  
at the protostar position (Figures~5$b$ and 5$c$). In contrast with NGC\,1333-IRAS~4B \citep{Wat07}, no mid-IR OH and H$_2$O lines are
seen. 
The presence of [\NeII] 
and [\FeII] lines from energy
levels above a few thousands Kelvin,
together with the bright and velocity-shifted [\OI] line detected by \textit{Herschel},  
suggests the presence of fast dissociative shocks close to the protostar 
\citep[\textit{e.g.,}][]{Hol89,Neu89}.
Note that the same mid-IR lines are readily detected in Herbig-Haro objects and outflow lobes
\citep[\textit{e.g.,}][]{Neu06,Mel08}.
As discussed  in Sect.~4, the observed intensities of
H$_2$, H$_2$O and CO rotational lines are more typical of slower velocity  shocks.

The large amount of gas and dust in embedded YSOs requires to apply extinction corrections to
retrieve corrected mid-IR line luminosities. 
In this work we have taken grain optical properties from Laor \& Draine \citep{Lao93}
and computed an extinction curve  $A(\lambda)/A_V$ from IR to submm wavelengths characterized 
by  $R_V$=$A_V$/$E$($B$-$V$)=5.5 (the ratio of visual extinction to reddening) and a conversion factor
from extinction to hydrogen column density of 1.4$\times$10$^{21}$\,cm$^{-2}$\,mag$^{-1}$.
These values have been previously used to correct photometric observations of embedded YSOs in the Serpens cloud \citep{Eva09}. 
Table~1 lists the uncorrected mid-IR line luminosities and those using an arbitrary correction of  $A_V$=150. 
In the latter case, even the [\OI]63\,$\mu$m line would be affected by extinction 
(the de-reddened line flux would be $\sim$2\,times higher) and the [\OI]63/145\,$\mu$m
 line flux ratio would increase to $\sim$20.

\begin{figure}[ht]
\begin{center}
\vspace{0.4cm}
\resizebox{\hsize}{!}{\includegraphics[angle=-90]{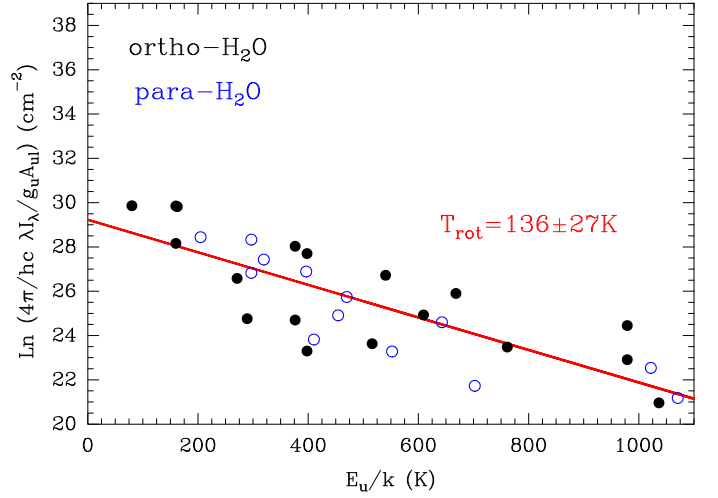}}
\caption{H$_2$O rotational diagram showing \textit{o}-H$_2$O (filled circles) and 
\textit{p}-H$_2$O far-IR  lines (open circles). 
Owing to subthermal excitation ($T_{\rm rot}$$\ll$$T_{\rm k}$) and large opacities of most lines, 
the H$_2$O diagram displays a larger   scatter (standard error on the fitted values are shown) than CO.}
\end{center}
\label{fig_h2orotdia}
\end{figure}

 Such large intensity ratios are predicted by dissociative shocks models
\citep[\textit{e.g.,}][]{Hol89,Neu89,Flo10} and thus provide a reasonable upper value to the extinction in the inner envelope.
An independent estimate of the extinction is obtained from  the $\sim$9.7\,$\mu$m  silicate grains
absorption band
 seen in the \textit{Spitzer} IRS low resolution spectra
\citep[shown in][]{Eno09}. In the diffuse ISM, the optical depth of the 
$\sim$9.7\,$\mu$m broad feature is proportional to $A_V$ 
\citep[\textit{e.g.,}][]{Whi03}.
Although it is not clear that the same correlation holds in the dense protostellar medium \citep{Chi07},
we computed a lower limit to the extinction of $A_V$$\gtrsim$30
by fitting the $\sim$9.7\,$\mu$m absorption band and assuming a single slab geometry.

\subsection{Dust Continuum Emission}

The dust continuum emission  peaks in the far-IR domain at $\sim$100\,$\mu$m. 
This is consistent with the presence of a massive dusty envelope  with relatively warm dust temperatures.
Using a simple modified blackbody  with a dust opacity varying as 
$\tau_{\lambda}$=$\tau_{100}$(100/$\lambda$)$^{\beta}$,
where $\tau_{100}$ is the continuum opacity at 100\,$\mu$m and $\beta$ is the
dust spectral index, we obtain a size of $\sim$7$''$ for the optically thick far-IR continuum source, with
$\tau_{100}$$\simeq$2.5, $\beta$=1.7, $L_{\rm far-IR}$$\simeq$26\,L$_{\odot}$ and a dust temperature of $T_{\rm d}$$\simeq$33\,K 
(Fig.~2).
These parameters agree with more sophisticated radiative transfer models
of the continuum emission detected with ISO/LWS,
for which a far-IR source size of $\sim$4$''$  ($\sim$500\,AU in radius) and $T_{\rm d}$=43\,K at 
the $\tau_{100}$=1 equivalent radius were
inferred \citep{Lar00}. 
They are also consistent with more recent SED models \citep{Kri12} 
and with the compact ($\sim$5$''$) emission 
revealed by mm interferometric observations and thought to arise from  the densest regions of the
inner envelope
and from outflow material \citep{Hog99, Eno09}.
Note that owing to the large continuum opacity below 100\,$\mu$m, observations at these
wavelengths probably do not trace the  innermost  regions, \textit{i.e.,}~the developing circumstellar disk with 
an approximate radius of $\sim$100\,AU \citep[\textit{e.g.,}][]{Rod05}.

\vspace{-0.2cm}
\section{Analysis}

\subsection{{\rm $^{12}$CO}, {\rm $^{13}$CO}, {\rm H$_2$O} and  {\rm OH} Rotational Ladders}

Figure~6 shows all $^{12}$CO and $^{13}$CO detected lines %(for which we could obtain accurate intensity measurements)
in a \textit{rotational diagram}. In this plot we assumed that the line emission arises from a source 
with a radius of 500\,AU (see Sect.~3.3).
Given the high densities  in  the inner envelope of protostars ($>$10$^6$\,cm$^{-3}$; see below), 
the rotational temperatures ($T_{\rm rot}$)
derived from these plots are a good lower limit to $T_{\rm k}$.
The $^{12}$CO diagram  suggests the presence of 3 different $T_{\rm rot}$ components 
with $T_{\rm rot}$$=$622$\pm$30\,K  (for $J_u$$\lesssim$42), $T_{\rm rot}$$=$337$\pm$40\,K 
($J_u$$\lesssim$26) and $T_{\rm rot}$$=$103$\pm$15\,K  ($J_u$$\lesssim$14) respectively. 
The estimated error in $T_{\rm rot}$ corresponds to different choices of $J_u$ cutoffs for the different
components\footnote{Note that an extinction correction of A$_V$=150  will only imply slightly larger far-IR line fluxes 
(more than a factor 1.5 below $\sim$80\,$\mu$m).
Slightly higher rotational temperatures would be inferred if A$_V$=150
($\sim$350\,K and $\sim$700\,K for the \textit{warm} and \textit{hot} components 
respectively).}.
 In the following we shall refer to them as the \textit{hot}, \textit{warm} and \textit{cool}
components.
Whether or not these $T_{\rm rot}$ components 
are associated with 3 real physical components
or with a more continuous temperature and mass distribution will be discussed in Section~4.1.
Here we  note that the  submm low-$J$ CO emission  shows broad line-profiles and a more extended 
distribution \citep{Dav99,Dio10} than the high-$J$  CO and H$_2$O lines detected with PACS 
(more sharply peaked near the protostar). 
In addition, velocity-resolved observations with HIFI of high-$J$ CO lines up to $J$=16-15 show 
different profiles compared with low-$J$ CO lines \citep[][Kristensen et al., in prep]{Yil12}. Specifically, for
the $^{12}$CO $J$=10-9 profile toward SMM1, Y{\i}ld{\i}z et al. (in prep.) find that about 1/3 of the integrated 
intensity is due to a narrow (FWHM of a few km\,s$^{-1}$) component originating from the quiescent envelope
and 2/3 to the broad outflow component. Thus, the cool gas component seen in the SPIRE submm maps is 
dominated by the entrained outflow gas. On the other hand, the CO $J$=16-15 profile of SMM1 is less broad
and more
similar to the excited H$_2$O line profiles observed with HIFI (Kristensen et al., in prep.). 
Thus, the far-IR CO and H$_2$O lines detected by PACS clearly probe different physical components than the
SPIRE data.

 \begin{table}[t]
      \caption[]{Observed and modelled luminosities.}
\centering
\begin{tabular}{cccc}
\hline\hline
\multicolumn{1}{c}{Species} & 
\multicolumn{1}{c}{Log$_{10}\,L_{obs}\,(L_\odot)$} & 
\multicolumn{1}{c}{$\%$$^a$} & 
\multicolumn{1}{c}{$\%$ of $L_{obs}$ in each component$^b$}  \\ \hline\hline
$^{12}$CO & $-$1.17 &  54\%  &$\sim$15\% (\textit{h}) $\sim$60\% (\textit{w}) $\sim$25\% (\textit{c})   \\
H$_2$O    & $-$1.56 &  22\%  &$\sim$90\% (\textit{h})\,\,\, $\sim$10\% (\textit{c})   \\
$[$\OI$]$ & $-$1.82 &  12\%  &$\sim$100\%   (\textit{w})    \\
OH        & $-$1.94 &  9\%   &$\sim$100\%   (\textit{w})   \\
$^{13}$CO & $-$2.75 &  1\%   &$\sim$40\% (\textit{w})\,\,\, $\sim$60\% (\textit{c})   \\
$[$\CII$]$& $-$3.32 &  0\%   &$\sim$100\% (\textit{w})  \\
$[$\CI$]$ & $-$3.42 &  0\%   &$\lesssim$10\% (\textit{w}) $\gtrsim$90\% (\textit{c})  \\%$\sim$100\%   (\textit{w})   \\
Dust & 1.41   & - & \\\hline
H$_2$ & $-$(4.83-2.54$^\dagger$)  &  & \\
$[$\FeII$]$ & $-$(3.57-2.62$^\dagger$) &   & \\
$[$\SiII$]$ & $-$(3.98-2.03$^\dagger$) &   & \\
$[$\SI$]$ &   $-$(4.52-3.07$^\dagger$) &   & \\
$[$\NeII$]$ & $-$(4.93-2.67$^\dagger$) &   & \\\hline
\end{tabular}
\tablefoottext{a}{Fraction of total far-IR/submm line luminosity.\\}
\tablefoottext{b}{From simple model:~\textit{h}=\textit{hot}, \textit{w}=\textit{warm},
 \textit{c}=\textit{cool} component.\\}
\tablefoottext{\dagger}{From \textit{Spitzer} IRS observations (10-37\,$\mu$m). 
Upper limits include an arbitrary extinction correction of A$_{V}$=150\,mag.}
\end{table}

Alternatively, Neufeld (2012) pointed out that the  shape of the rotational diagrams do not, by themselves, 
necessarily require multiple temperature components. 
In particular, a
single-$T_{\rm rot}$ solution  could be found for the    $^{12}$CO $J_u$$>$14 lines
in Serpens\,SMM1  if the gas were very hot ($T_{\rm kin}$$\simeq$2500\,K) but had 
a low density (a few~10$^4$\,cm$^{-3}$). This scenario seems less likely, at least for 
the circumstellar gas in the vicinity of the protostar. Note that the gas density in the inner envelope
of SMM1 is necessarily higher, as probed by our detection of high-$J$  lines from high dipole moment 
molecules such as HCN and HCO$^+$ (with very high critical densities). 
In addition, detailed SED models
predict densities
of $n$(H$_2$)$\simeq$4$\times$10$^6$\,cm$^{-3}$ at $\sim$1000\,AU, whereas
densities of the order of $\sim$10$^5$\,cm$^{-3}$ are only expected at much larger radii 
\citep{Kri12}.
Therefore, it seems more likely that the different $T_{\rm rot}$ slopes probe different
physical components or the presence of a  temperature  gradient. %in the same component.

A  rotational diagram of the detected $^{13}$CO lines provides 
$T_{\rm rot}$=76$\pm$6\,K, thus  lower than the $T_{\rm rot}$ inferred for
 $^{12}$CO in the \textit{cool} component. The measured  $^{12}$CO/$^{13}$CO $J$=5-4 
line intensity ratio is $\simeq$8, thus much lower than the typical $^{12}$C/$^{13}$C isotopic ratio of $\sim$60
\cite{Lan90}, whereas the  $^{12}$CO/$^{13}$CO $J$=16-15 intensity ratio  is $\simeq$55.
These different ratios show that the submm $^{12}$CO low- and mid-$J$    lines are optically thick, 
but the high-$J$ far-IR   lines are optically thin.

An equivalent plot of the observed far-IR H$_2$O line intensities   
in a rotational diagram  gives $T_{\rm rot}$=136$\pm$27\,K without much indication of multiple
$T_{\rm rot}$ components (Figure~7).
Since H$_2$O critical densities are much higher
than those of CO, collisional  thermalization only occurs at very high densities ($>$10$^8$\,cm$^{-3}$).
Therefore, the large scatter of the  H$_2$O rotational diagram is a consequence of 
subthermal excitation conditions ($T_{\rm rot}$$\ll$$T_{\rm k}$) and large 
H$_2$O line opacities. 
We anticipate here that the opacity of most (but not all) far-IR H$_2$O lines is very high 
($\tau$$\gg$1). 
Compared to CO, the H$_2$O rotational diagram is thus obviously less meaningful.
A similar rotational diagram of the observed OH lines provides $T_{\rm rot}$=72$\pm$8\,K 
(see also Wampfler et al., submitted), even lower than the inferred H$_2$O rotational temperature.
Like H$_2$O, OH transitions also have high critical densities and line opacities 
(except the high-energy and the cross-ladder transitions).

\vspace{-0.2cm}
\subsection{Simple Model of the Far-IR and Submm Line Emission}

Observations and models of the protostellar environment suggest 
that the observed emission can arise from different physical components, where different 
mechanisms dominate the gas heating and  the prevailing chemistry: hot and warm gas from energetic shocks
in the inner envelope
(\textit{e.g.,}~small scale shocks along the outflow cavity walls, bow shocks and 
working surfaces within an atomic jet);
UV-illuminated gas in the cavity walls; cooler gas from the envelope passively heated
by the protostar luminosity  and
the entrained outflow gas  \citep[\textit{e.g.,}][]{vD11}. 
Unfortunately, the lack of enough spectral and angular resolution of our data
does not allow us to provide a complete characterization of all the different possible components. 
However, the large number of detected 
lines, high excitation ($E_{\rm u}$/$k$$>$300\,K) optically thin lines in particular, 
helps to determine the dominant contributions and the average physical conditions.

With this purpose, we have carried out
a  non-LTE radiative transfer model  of the central spaxel far-IR and submm spectrum (the protostar position) 
using a multi-molecule LVG code \citep{Cer12}. 
 In this model we included three spherical components 
suggested by the three temperature components seen in the $^{12}$CO rotational diagram
(\textit{i.e.,} \textit{hot}, \textit{warm} and \textit{cool} components).
In our model, CO is present in the three components, but since a single rotational temperature roughly 
explains the observed far-IR H$_2$O and OH lines, most of their emission in the PACS domain can be reproduced
with 
only one component, the \textit{hot} component for H$_2$O and the \textit{warm} component for OH (see next section
for more details).
Note that we do not model the extended quiescent envelope with the bulk of the (cold) mass seen in
the mm continuum and narrow C$^{18}$O line emission (Y{\i}ld{\i}z et al. in prep.).
Foreground absorption and emission from the low density and low temperature extended envelope are 
thus not included in the model but they
will basically influence narrow velocity ranges of the lowest energy-level lines. 

The latest available collisional rates were used (\textit{e.g.,}~Daniel et al. 2011 and references therein for 
H$_2$O and Yang et al. 2010, extended by Neufeld 2012, for CO).
Although only one line of $o$-H$_2$ and one line of $p$-H$_2$
are detected  towards the Serpens~SMM1 protostar position, the large H$_2$~0-0 $S$(2)/$S$(1) line intensity ratio suggests
a H$_2$ \textit{ortho}-to-\textit{para}  (OTP) ratio lower than 3 in the gas probed by these low excitation H$_2$ lines.
Here we shall adopt an OTP ratio for H$_2$   of 1 (the value we obtain
from the two observed H$_2$ lines assuming LTE and a rotational temperature of $\sim$800\,K).
Note that such low non-equilibrium H$_2$  OTP values have been inferred, for example, in the hot shocked gas 
towards HH54 and HH7-11 \citep{Neu06}. 
The water vapour OTP ratio in the model, however, is let as a free parameter.

\begin{table}[t]
      \caption[]{Model components and source-averaged column densities 
                 (assuming a line-width of 20\,km\,s$^{-1}$ in all components).}
\centering
\begin{tabular}{llll}
\hline\hline
\multicolumn{1}{l}{Comp.} & 
\multicolumn{1}{l}{Radius} &
\multicolumn{1}{l}{$T_{\rm k}$ (K)} &
\multicolumn{1}{l}{$n$(H$_2$) (cm$^{-3}$)}\\\hline\hline
 \textit{Hot} &   $\sim$500\,AU & 800 & 5$\times$10$^6$  \\\hline
$N$(cm$^{-2}$) &  $^{12}$CO(5$\times$10$^{16}$)      &  H$_2$O(2$\times$10$^{16}$)   &   \\
                &  $^{13}$CO(9$\times$10$^{14}$)    &     &     \\\hline\hline
\textit{Warm}   &   $\sim$500\,AU & 375 & 5$\times$10$^6$  \\\hline
$N$(cm$^{-2}$)  & $^{12}$CO(1$\times$10$^{18}$)    &  $^{13}$CO(2$\times$10$^{16}$) &  \\
                & OH(1$\times$10$^{16}$)           &  HCO$^+$(5$\times$10$^{14}$)   &  \\
                & HCN(2$\times$10$^{14}$)        &  CH$^+$($<$2$\times$10$^{14}$) &  \\
                & CO$^+$($<$2$\times$10$^{15}$)  &  NH$_3$($<$2$\times$10$^{15}$)  &  \\
                & O(1$\times$10$^{18}$)          &      &  \\\hline\hline
\textit{Cool}   &  $\sim$4500\,AU & 140 &  2$\times$10$^5$        \\\hline
$N$(cm$^{-2}$)  &  $^{12}$CO(1$\times$10$^{17}$)   &  H$_2$O(8$\times$10$^{15}$)  &  \\
                &  $^{13}$CO(1$\times$10$^{16}$)   &  HCO$^+$(2$\times$10$^{14}$) &  \\
                &  HCN(1$\times$10$^{15}$)         &  C(1$\times$10$^{17}$)       &  \\
                &  C$^+$(1$\times$10$^{16}$)       &                              &  \\\hline\hline

%H$_2$O/CO & $<$0.2     &   \textit{cool} \\\hline
\end{tabular}
\vspace{-0.4cm}
\end{table}

The $o$-H$_2$O ground-state line observed with HIFI towards Serpens\,SMM1 shows a two component emission 
line profile with a \textit{medium} component ($\Delta$v$\simeq$15\,km\,s$^{-1}$) thought to arise from 
small-scale shocks in the inner cavity walls and a \textit{broad} component 
($\Delta$v$\simeq$40\,km\,s$^{-1}$) 
from more extended shocked-gas along the outflow \citep{Kri12}.
In addition, velocity-resolved observations of the OH $^2\Pi_{1/2}$ $J$=3/2-1/2 line ($\sim$163.1$\mu$m)  
with HIFI
show a broad line-profile with $\Delta$v$\simeq$20\,km\,s$^{-1}$ (Wampfler, S. priv.~comm.).
For simplicity here we shall adopt a typical  line-width of 20\,km\,s$^{-1}$ in all modelled components.
As in any LVG calculation, the model is more accurate for (effectively) optically thin  lines 
(high-$J$ CO lines, excited and weak H$_2$O and OH lines, etc.) than for very opaque lines 
(\textit{e.g.,} low excitation lines).

\subsubsection{Model: Hot and Warm Component} 

Following Sect.~3.3. and previous far-IR studies of this source, we have assumed a size of $\sim$4$''$ 
\citep[$\sim$500\,AU in radius; \textit{e.g.,}][]{Lar00,Kri12} 
for the far-IR source (\textit{hot} and \textit{warm} components),
a mixture of  shocked and UV-illuminated gas (see Section~5).
A good fit to the  high-$J$ $^{12}$CO emission is
obtained for $n$(H$_2$)$\approx$5$\times$10$^{6}$\,cm$^{-3}$ 
 with $T_{\rm k}$$\simeq$800\,K and $\simeq$375\,K for the \textit{hot} and \textit{warm} components respectively.
The approximate mass in these components is only $M$$\lesssim$0.03\,$M$$_{\odot}$.
Although this solution is not unique,
this model satisfactorily reproduces  not only the  $^{12}$CO high-$J$ lines but also the lines from other species 
(note that we searched for a combination of $T_{\rm k}$ and $n$(H$_2$) that reproduces all species simultaneously; see below).

\begin{figure*}[t]
\begin{center}
\resizebox{\hsize}{!}{\includegraphics[angle=-90]{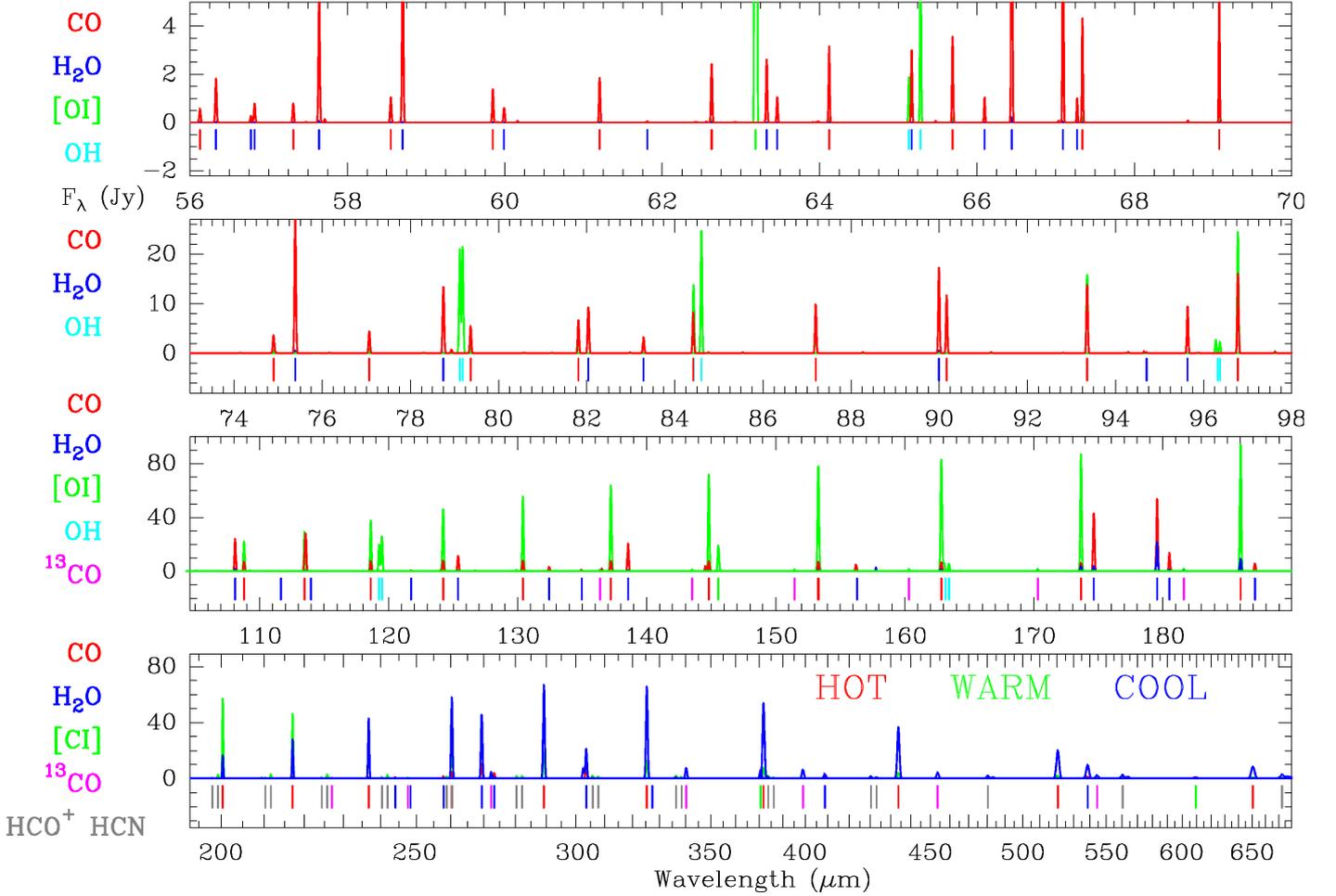}}
\caption{Synthetic non-LTE LVG spectrum of the  Class~0 protostar Serpens MM1 convolved to PACS  (\textit{top three panels}) 
and SPIRE (\textit{lower panel}) 
spectral resolutions. Continuous curves show the line emission
contribution from the ``\textit{hot}'' (red), ``\textit{warm}'' (green) and ``\textit{cool}'' (blue) components 
discussed in the text. Vertical labels mark the wavelength position of CO (red), H$_2$O (blue), OH (cyan), 
$^{13}$CO (magenta), HCO$^+$ and HCN (grey) rotational transitions. 
[O\,{\sc i}]63, 145 and [C\,{\sc i}]370, 609\,$\mu$m fine structure lines are marked with green labels.} 
\end{center}
\label{fig_lvgcomps}
\end{figure*}

\begin{figure*}[t]
\begin{center} 
\resizebox{\hsize}{!}{\includegraphics[angle=-90]{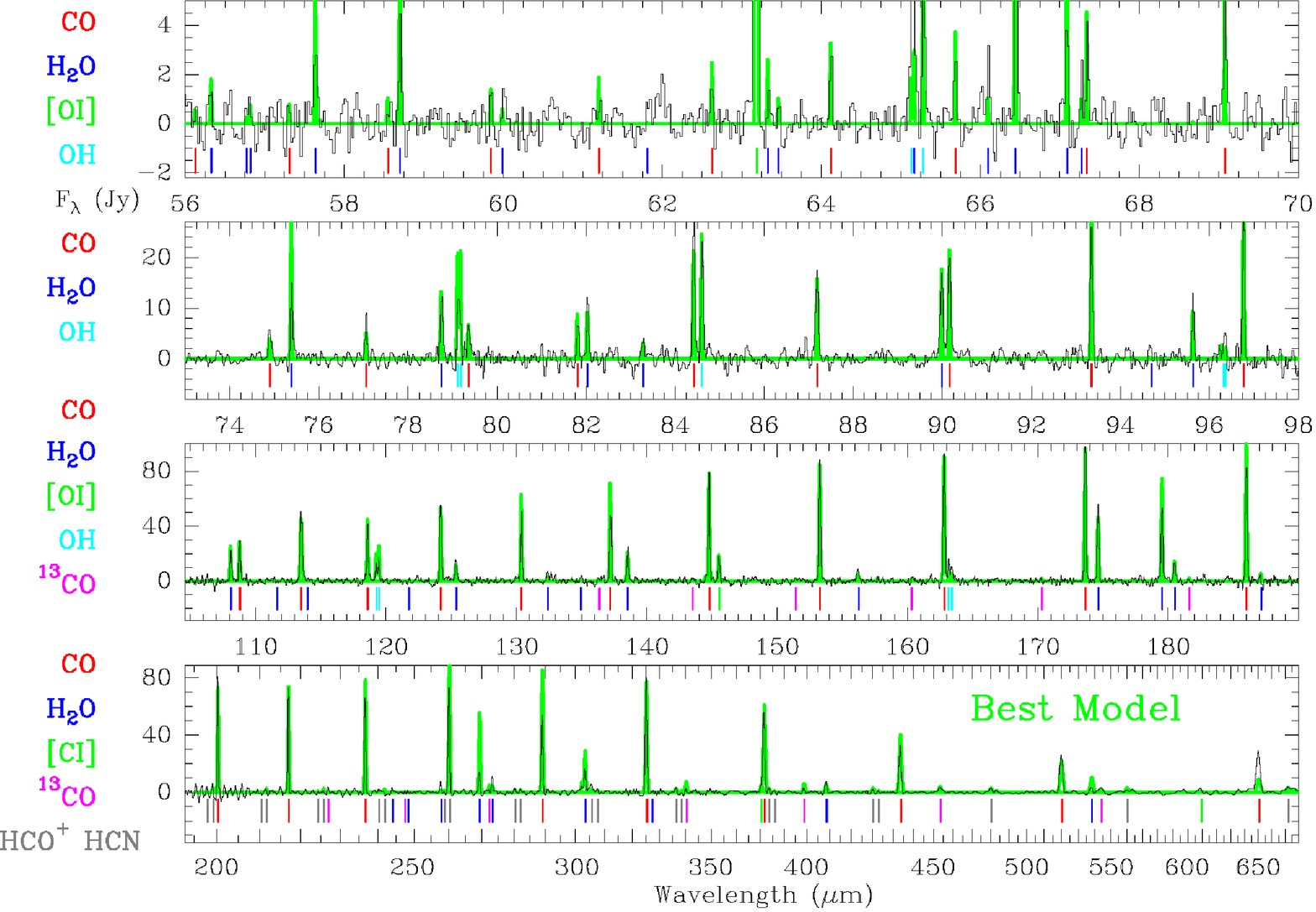}} 
\caption{\textit{Herschel} far-IR and submm spectrum of Serpens~SMM1 at the central spaxel position (black histogram) and complete model
(green continuous curves) adding the emission from the \textit{hot}, \textit{warm} and \textit{cool} components
and convolved to PACS and SPIRE spectral resolutions.} 
\end{center}
\label{fig_partmodel}
\end{figure*}

In order to constrain the density and place H$_2$O and OH in a particular model component, we first
run a grid of excitation models and investigated how  particular
line ratios change with $n$(H$_2$) and $T_{\rm k}$. In this comparison we choose faint, low-opacity lines
arising from high energy levels that are observed with a similar PSF. 
Figure~A.1  shows the variation of the $o$-H$_2$O 3$_{30}$-3$_{03}$/6$_{15}$-3$_{05}$ (67.269/82.031\,$\mu$m)
and $p$-H$_2$O 6$_{06}$-5$_{15}$/5$_{15}$-4$_{04}$ (83.284/95.627\,$\mu$m) line ratios.
The former ratio shows that the excited H$_2$O lines arise from dense gas, whereas
the latter requires high temperatures ($>$500\,K). The  intersection of the two line ratios
is consistent with densities higher  than a few 10$^6$\,cm$^{-3}$ and temperatures around $\sim$800\,K. %. 
Note that for $T_{\rm k}$$<$400\,K, the predicted
emission from many  excited H$_2$O lines below $\sim$100\,$\mu$m would be too faint.
 On the other hand, 
Figure~A.2 shows a very clear dependence of the OH 65.279/96.363\,$\mu$m line ratio    
with the gas temperature. The observed ratio does suggest that OH arises from gas at  $T_{\rm k}$$\lesssim$400\,K.
Following the previous excitation analysis, and  according to the higher observed rotational
temperatures ($T_{\rm rot}^{\rm H_2O}$$>$$T_{\rm rot}^{\rm OH}$) 
and higher H$_2$O energy levels detected, 
we placed H$_2$O in the \textit{hot} component
and OH (together with [\OI]) in the \textit{warm}  component. 
Although this choice provides a better fit to the data, we can not obviously conclude that the bulk of the water vapour and OH 
emission arise 
from different physical components.

Far-IR radiative pumping is not included in our models and thus 
the derived  H$_2$O and OH column densities might be considered as upper limits.
Nevertheless, all  H$_2$O and  OH  lines are observed in emission and show $T_{\rm rot}$$>$$T_{\rm d}$
(where $T_{\rm d}$ is the dust temperature inferred from the SED analysis, see Sect.~3.3),
suggesting that owing to the high densities but moderate far-IR radiation field  at $\gtrsim$500\,AU,  
collisions dominate  at first order. 
Note that high-mass protostars produce much stronger far-IR dust continuum fields (by orders of magnitude)
and many far-IR OH and  water vapour lines
are often observed in absorption or show P-Cygni profiles when far-IR pumping dominates the excitation
\citep[see \textit{e.g.,}][for far-IR OH and H$_2$O  lines in Orion~KL outflows]{Goi06,Cer06}.

\subsubsection{Model: Cool Gas Component} 

In order to fit the more extended  $^{12}$CO low- and mid-$J$  emission 
and to match the $T_{\rm rot}$$\simeq$100\,K component 
seen in the $^{12}$CO rotational diagram (Figure~6),   
we included a third \textit{cool} component --~the entrained outflow gas.
From the SPIRE $^{12}$CO maps we infer an approximate radius of $\sim$20$''$ ($\sim$4500\,AU) 
for the emitting region.
For this geometry, a density of  $n$(H$_2$)$\approx$2$\times$10$^{5}$\,cm$^{-3}$ and a temperature
of  $T_{\rm k}$$\simeq$140\,K produces a good fit of the observed $^{12}$CO submm emission 
\citep[see also][for velocity-resolved observations in NGC 1333 IRAS 4A/4B]{Yil12}.
This component is also needed to fit the lower excitation H$_2$O  submm  lines.  

Low- and mid-$J$ $^{13}$CO lines are optically thin and thus, in addition to the swept-up or entrained
outflow gas, velocity-resolved submm $^{13}$CO line profiles  
would carry information about other components such as the UV-heated gas
and the quiescent envelope, which may well dominate the $^{13}$CO lower-$J$ emission \citep{Yil12}. 
Detailed models of the  emission from the  extended and
 passively heated envelope (having most of the mass) 
predict  $^{12}$CO rotational temperatures around $\simeq$30-60\,K  \citep[][Harsono et al., in prep]{Vis12}.

To summarize, Table~2 shows the model parameters and 
Figure~8 shows the entire $\sim$55 to 671\,$\mu$m  synthetic spectrum for the \textit{hot} (red),  
\textit{warm} (green) and \textit{cool} (blue) 
components convolved with the PACS and SPIRE resolutions. 
A comparison of the resultant synthetic spectrum (by simply adding the 3 component spectra)
with the \textit{Herschel} spectra is shown in Figure~9.

\vspace{-0.3cm}
\subsubsection{Columns, Abundances and Validity of the Model}

Owing to the lack of angular resolution to determine the beam filling factors of the
different physical components towards Serpens~SMM1, the relative abundance ratios derived from the model 
and shown in Table~3 are a better diagnostic tool
than the absolute column densities.
Nevertheless, here we provide   the source-averaged column densities ($N$)
in the different components as well as the upper limits for several non-detected species (Table~2). 
In the  \textit{hot} component we find
$\sim$5$\times$10$^{16}$\,cm$^{-2}$ and $\sim$2$\times$10$^{16}$\,cm$^{-2}$
column densities for $^{12}$CO and H$_2$O  respectively. Note that a water vapour OTP ratio of  
$\sim$3 provides the best fit to the observed water vapour lines and this is the value adopted in the models.
Assuming that the gas in the \textit{warm} component ($T_{\rm k}$$\simeq$375\,K)
covers a similar area,
we obtain $N$(OH)$\sim$10$^{16}$\,cm$^{-2}$ and $N$($^{12}$CO)$\sim$10$^{18}$\,cm$^{-2}$ 
(a factor $\approx$20 larger than $N$($^{12}$CO) in the \textit{hot} component).

The detected mid-IR H$_2$ lines provide a lower limit  to the H$_2$ column density of 
$N_{\rm H_2}$$\approx$10$^{22}$\,cm$^{-2}$. We use
this column density to provide an upper limit
to the absolute  abundances (with respect to H$_2$) in the \textit{hot}+\textit{warm}
 components. In particular we obtain $\lesssim$10$^{-4}$,
 $\lesssim$0.2$\times$10$^{-5}$ and $\lesssim$10$^{-6}$ for the CO, H$_2$O  and OH abundances respectively. 
The inferred upper limit to the water vapour abundance 
 is much higher than the $\approx$10$^{-(8-9)}$  value typically found in cold interstellar clouds 
\citep[\textit{e.g.,}][]{Cas10}  but is lower than the
$\approx$10$^{-4}$ value often expected in the hot shocked gas \citep[\textit{e.g.,}][]{Kau96}.
Note that the high source-averaged column density of atomic 
oxygen ($\sim$10$^{18}$\,cm$^{-2}$) 
 suggests that a significant fraction of gas-phase oxygen reservoir in shocks is kept in atomic form. 

The column densities in the entrained outflow gas  (\textit{cool} component at 
$T_{\rm k}$$\simeq$140\,K) are given in Table~2. The absolute columns are more uncertain
as they depend on the assumed spatial distribution of the swept-up outflow gas.
\begin{table}[t]
      \caption[]{Selected abundance ratios in the  modelled components.}
\centering
\begin{tabular}{ccc}
\hline\hline
\multicolumn{1}{c}{Species} & 
\multicolumn{1}{c}{Abundance ratio} &
\multicolumn{1}{c}{Component}\\\hline\hline
H$_2$O$_h$/(CO$_w$+CO$_h$)  & $\sim$0.02 & \textit{hot} and \textit{warm}  \\
OH$_w$/(CO$_w$+CO$_h$)      & $\sim$0.01 & \textit{hot} and \textit{warm}  \\
O$_w$/(CO$_w$+CO$_h$)       & $\sim$0.8    & \textit{hot} and \textit{warm}  \\
OH$_w$/H$_2$O$_h$           & $\lesssim$0.5  & \textit{warm}/\textit{hot} \\
CO$_w$/CO$_h$               & $\sim$20   & \textit{warm}/\textit{hot} \\
H$_2$O$_h$/CO$_h$           & $\sim$0.4  & \textit{hot} \\
CH$^{+}_{h}$/H$_2$O$_h$       & $<$0.01    & \textit{hot}   \\
O$_w$/OH$_w$                & $\sim$100  & \textit{warm}  \\
HCO$^{+}_{w}$/HCN$_w$       & $\sim$2.5  & \textit{warm}  \\
HCO$^{+}_{c}$/HCN$_c$       & $\sim$0.2  & \textit{cool} \\
H$_2$O$_c$/CO$_c$           & $<$0.08    & \textit{cool} \\\hline
\end{tabular}
%\vspace{-0.5cm}
\end{table}
Note that in order to fit the low-$J$ $^{13}$CO lines in the  \textit{cool} component, we had to include
a low $^{12}$CO/$^{13}$CO column density ratio of $\simeq$10, confirming
that the low-J $^{12}$CO lines are optically thick and thus they do not probe the bulk of the material
seen in the low-J CO isotopologue line emission (\textit{i.e.,} the massive and quiescent envelope).

Our simple LVG model satisfactorily reproduces the absolute fluxes of most observed  lines and does not predict
lines that are not detected in the spectra (see Figure~9).
Despite the fact that this model solution is obviously not unique, the agreement with observations suggests
that this model 
captures the average physical conditions of the shocked gas near the protostar. The level of agreement is
typically better than $\sim$30\% (\textit{i.e.,}~similar to the calibration uncertainty). 
The worst agreement  occurs for several low-excitation 
optically thick lines that can arise from different physical components because 
the non-local radiative coupling between different components is not treated in the LVG model. 
In addition, some lines such as the $\sim$163\,$\mu$m OH lines 
\citep[sensitive to far-IR pumping, see][]{Off92,Goi02} are underestimated by $\gtrsim$40\%, suggesting that radiative pumping may
play some role (see detailed OH models by Wampfler et al., submitted).

Regarding the $T_{\rm k}$ and $n$(H$_2$) conditions inferred in each component, Figure~A.3 
in the Appendix shows that for a given
gas density, we can  distinguish temperature variations of $\sim$30\%  (especially in high-$J$ CO and
in some high excitation H$_2$O and OH lines). Figure~A.4 shows that for a given gas temperature, 
density variations of a factor $\sim$2-3 can also be distinguished.
Therefore, for the assumed geometry and physical conditions,
the derived source-averaged  column densities are accurate within a factor of $\sim$2.

\vspace{-0.2cm}
\section{Discussion}

\subsection{Shocked-Gas Components}

Several fine-structure lines that probe the very hot  atomic gas (a few thousand \,K)
 near the protostar  are detected in the mid-IR.
Besides, most tracers of the shock-heated  molecular gas  (a few hundred\,K) appear in the far--IR.
At longer submm and mm wavelengths, extended emission from cool  entrained outflow gas 
and from the cold massive envelope dominates.  
The relative intensities of the detected mid- and far-IR atomic and molecular lines
help to qualitatively constrain the nature of the main shocked-gas components
in Serpens\,SMM1. 

The shock wave velocity, pre-shock density and the magnetic field strength determine most 
of the shocked gas properties.
Fast shocks can destroy molecules and ionize atoms, whereas slower shocks heat the gas
without destroying molecules. Depending on the evolution of the shock structure it is common to distinguish
between $J$-type (or Jump) and $C$-type (or Continuous). More complicated, ``mixed'' non-stationary situations
may also exist \cite[see reviews by \textit{e.g.,}][]{Dra93,Wal05}. 
As we show below, our observations suggest the presence of both fast and slow
 shocks in Serpens\,SMM1. 

\subsubsection{Fast Shocks}

The bright and velocity-shifted [\OI]63\,$\mu$m emission, together with the detection of 
[\NeII]12\,$\mu$m and very high energy [\FeII] lines in Serpens\,SMM1 suggests the presence of fast
dissociative $J$-shocks related with the presence of an embedded  atomic jet near the protostar. 
Note that [\NeII] and [\FeII] lines have been detected towards other  YSOs  \citep[see \textit{e.g.,}][for the c2d \textit{Spitzer} sample]{Lah10}.
Hollenbach \& McKee (1989) presented detailed models for the fine-structure emission
of atoms and ions in such dissociative $J$-type shocks. However, the chemistry of S, Fe, Si and related molecules 
\citep[including gas-phase depletion, \textit{e.g.,}][]{Nis05} were not included and thus the fine-structure absolute
line intensity predictions  are likely upper limits.
In addition, because the beam filling factors of the  possible
shock components are not known, line ratios provide a much better diagnostic than absolute intensities.
According to these detailed models, the observed [\SiII]35\,$\mu$m/[\FeII]26\,$\mu$m~$<$1 %and [\SiII]35\,$\mu$m/[\SI]25\,$\mu$m~$<$1 
line intensity ratio (independently on the assumed extinction)
provides a lower limit to the pre-shock density, $n$$_{\rm 0}$=$n_0$(H)+2$n_0$(H$_2$),  of $>$10$^{4}$\,cm$^{-3}$.
Besides, owing to the high ionization potential of Ne$^+$, the [\NeII]12\,$\mu$m/[\FeII]26\,$\mu$m line
intensity ratio increases sharply with the shock velocity (v$_{\rm s}$).
A lower limit of v$_{\rm s}$$>$60\,km\,s$^{-1}$ is found from the
observed [\NeII]12\,$\mu$m/[\FeII]26\,$\mu$m~$\geq$0.1
line ratio (applying an extinction correction of A$_V$$\geq$30).

Models of fast dissociative $J$-shocks predict that the gas    is initially atomic 
and very hot (a few thousand\,K), but by the time that molecules reform, the gas cools to about 400\,K.
H$_2$ formation provides the main heat source for this ``temperature plateau''
\citep{Hol89,Neu89}. In such models, 
H$_2$ is not so abundant and cooling by H$_2$ and H$_2$O can be less important than by [\OI], CO and OH.
Therefore, in addition to the mid--IR fine-structure emission,
a significant fraction of the [\OI], OH and  the CO \textit{warm} component 
emission  
can arise behind fast dissociative shocks triggered by a jet impacting the inner, dense evelope. 

\subsubsection{(UV-irradiated) Slow Shocks}

From the observed H$_2$ 0-0 $S$(2) weak line and the upper limit of the non-detected $S$(0) line, we derive a lower limit to the
H$_2$ rotational temperature ($T_{\rm 42}$) of $\sim$700\,K 
(if a extinction correction of A$_V$=30 is applied). 
This (and higher) temperatures of the molecular gas
 are often inferred behind non-dissociative shocks  
\citep[\textit{e.g.,}][]{Neu06,Mel08}.

Depending on  the shock velocity, non-dissociative $C$-type shocks shielded from UV radiation 
 can produce very high gas temperatures  without destroying molecules 
($T_{\rm k}$$\sim$400-3000\,K for v$_{\rm s}$$\sim$10-40\,km\,s$^{-1}$ in the models by Kaufman \& Neufeld 1996). 
They also naturally produce high H$_2$O/CO abundance ratios (close~to~1) but predict low
OH/H$_2$O$\ll$1 ratios (owing to negligible H$_2$O photodissociation but very efficient
OH + H$_2$ $\rightarrow$ H$_2$O + H reactions in the hot molecular gas).
Therefore, a non-dissociative $C$-shock could be the origin of  the observed H$_2$ lines, 
high-$J$ CO ($J_{\rm u}$$\gtrsim$30) 
and excited H$_2$O lines ($E_{\rm u}$/$k$$>$300\,K). 
Note that in our simple model we infer a high H$_2$O$_h$/CO$_h$$\simeq$0.4 relative abundance 
when considering  the \textit{hot} 
component alone. 
On the other hand, if the \textit{hot} and \textit{warm} components inferred from the $^{12}$CO rotational diagram
only reflect the presence of a temperature gradient in the same physical component,  
then we would infer much lower 
H$_2$O$_h$/(CO$_h$+CO$_w$)$\simeq$0.02 abundances (see Table~3). 
Such low H$_2$O abundances are  difficult to reconcile with non-dissociative $C$-shocks
shielded from UV radiation and are more consistent with low-velocity $J$-shocks (see below).

\begin{figure}[t]
\begin{center}
\resizebox{\hsize}{!}{\includegraphics[angle=-90]{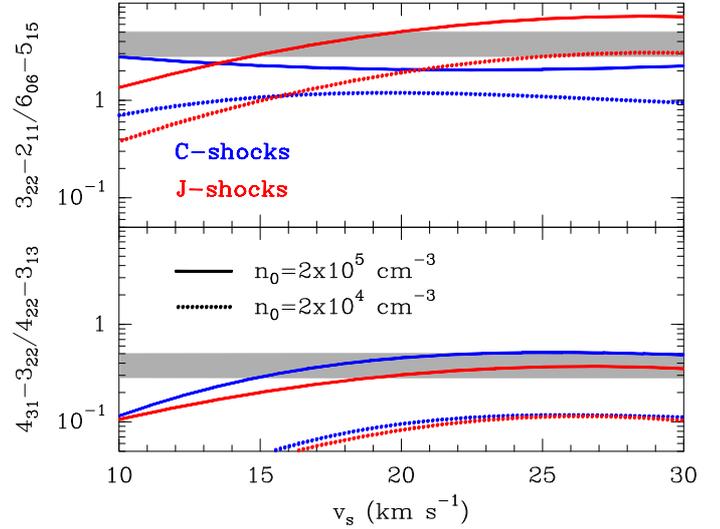}}
\caption{Comparison of selected  $p$-H$_2$O line intensity ratios observed with PACS 
 and shock model predictions from Flower \& Pineau Des For{\^e}ts (2010).
Red and blue curves represent $J$-type and $C$-type shock models respectively.
The pre-shock gas density is $n_0$=2$\times$10$^5$\,cm$^{-3}$ (continuous curves) and 
$n_0$=2$\times$10$^4$\,cm$^{-3}$ (dotted curves).
Low opacity lines from high energy rotational levels ($E_{\rm u}/k$$\sim$300-650\,K) observed at similar
wavelengths are selected. 
The \textit{upper} panel shows the $p$-H$_2$O 3$_{22}$-2$_{11}$/6$_{06}$-5$_{15}$ line ratio (89.988/83.284\,$\mu$m)
and the \textit{lower} panel shows the 4$_{31}$-3$_{22}$/4$_{22}$-3$_{13}$ (56.325/57.637\,$\mu$m) line ratios. 
The grey horizontal areas show the observed ratios and their error margins.}
\end{center}
\label{fig_shockmods}
\vspace{-0.5cm}
\end{figure}

Planar-shock model predictions of absolute H$_2$O line intensities 
depend again on the unknown beam filling factor of the shocked material. Figure~10 shows a
comparison of selected H$_2$O line intensity ratios   
with recent unidimensional models of non-dissociative  $C$-type and $J$-type shocks 
(the latter for v$_{\rm s}$$<$30\,km\,s$^{-1}$) 
by Flower \& Pineau Des For{\^e}ts (2010).
In order to avoid using optically thick line diagnostics, different beam dilutions and extinction corrections,
we selected low opacity $p$-H$_2$O lines arising from 
high energy levels ($E_{\rm u}/k$$\sim$300-650\,K) observed at similar
wavelengths.
The \textit{upper} panel shows the 3$_{22}$-2$_{11}$/6$_{06}$-5$_{15}$ line ratio (89.988/83.284\,$\mu$m)
and the \textit{lower} panel shows the 4$_{31}$-3$_{22}$/4$_{22}$-3$_{13}$ (56.325/57.637\,$\mu$m) line ratio. 
The grey horizontal areas show the observed line intensity ratios and their error margins.
From this comparison we conclude that low-velocity, non-dissociative shocks (v$_{\rm s}$$\lesssim$20\,km\,s$^{-1}$) with 
a pre-shock density of $n_0$$\simeq$2$\times$10$^5$\,cm$^{-3}$ 
are needed to reproduce the observed ratios. 
Figure~10 favors  small-scale non-dissociative $J$-shocks, although  
it is difficult to determine any possible contribution of  $C$-type shocks in the frame
of current shock models and lack of spectral/angular resolution at $\lambda$$<$100\,$\mu$m.

Besides, the very different shock velocities required by Ne$^+$ versus H$_2$O lines suggests
that both emissions arise from different locations. This will be the case if the
Ne$^+$ emission arises from fast shocks within the jet itself, but the H$_2$O (and H$_2$) emission arises in lower
velocity shocks along the inner walls of the outflow cavity.  %\\\\
Finally, note that the similar excitation conditions in the hot and dense gas
($P/k$$=$ $n$\,T$_{\rm K}$$\simeq$8$\times$10$^9$\,K\,cm$^{-3}$),
as well as the similar  spatial distribution of the nearby
 CO $J$=29-28 and $p$-H$_2$O 3$_{22}$-2$_{11}$ lines shown in Figure~4
suggests that the  \textit{hot} CO and the excited H$_2$O lines arise in the same shock component.

The UV radiation field probed by the [\CII]158\,$\mu$m line can also heat the 
gas along the outflow cavity walls
\citep[see \textit{e.g.,}][]{Vis12}. UV photons will modify the chemistry of the shocked gas
(\textit{e.g.,} by photodissociating H$_2$O and enhancing the OH and O abundances; see Sect.~5.3), but in terms of gas heating,
 they are likely  not  a dominant heating mechanism of the hot gas emission seen towards
low-mass  protostars (see Karska et al. submitted).
All in all, the relative line intensities of the detected atomic and molecular species 
 confirm the presence of both fast dissociative and lower velocity, non-dissociative 
shocks that are irradiated by UV radiation fields.

\subsection{Line Cooling}

Complete far-IR and submm spectral scans allow an unbiased determination of the line cooling in YSOs.
Line photons following collisional excitation and escaping the region are
responsible of the gas cooling. Hence, the observed line luminosities in a broadband spectral scan provide a good measurement of the
total gas cooling budget (see Table~1).
In Serpens\,SMM1, $\sim$54\% of the total line
luminosity is due to $^{12}$CO lines, followed by   
H$_2$O ($\sim$22\% and $L$(H$_2$O)/$L$($^{12}$CO)$\simeq$0.4), [\OI] ($\sim$12\%) and OH (9\%). 
The total far-IR and submm line luminosity   
$L_{\rm FIR\,SMM}$=$L$($^{12}$CO)+$L$(H$_2$O)+$L$(\OI)+$L$(OH)+$L$($^{13}$CO)+$L$(C$^+$) is $L_{\rm FIR\,SMM}$$\simeq$0.12\,L$_{\odot}$, and the ratio
 $L_{\rm mol}$/$L_{\rm bol}$
 between the molecular line luminosity, $L_{\rm mol}$=$L$($^{12}$CO)+$L$(H$_2$O)+$L$(OH),
and the bolometric luminosity is $\simeq$3.5$\times$10$^{-3}$, all consistent with the expected emission of a Class~0 source
\citep[see \textit{e.g.,}][]{Gia01}.
Note that only  if the extinction in the mid-IR is larger than $A_V$$\simeq$200, then the intrinsic luminosity
of the observed H$_2$ lines will be higher than those of CO and H$_2$O.

In our model,  CO dominates the  
line cooling (Table~1) with more than half of the total CO luminosity arising from the \textit{warm} component
(a mixture of shocks and UV-illuminated material) and with a \textit{warm}/\textit{hot} luminosity ratio
of $L$(CO$_w$)/$L$(CO$_h$)$\simeq$5.
H$_2$O is the second most important species, with a dominant contribution
to the  hot shocked gas cooling.
Note that we predict that $\gtrsim$80\% of the H$_2$O line luminosity is radiated in the 
 55\,$\mu$m$<$$\lambda$$<$200\,$\mu$m range and only $\sim$5\%~at shorter wavelengths (in agreement with 
the absence of strong H$_2$O lines in the \textit{Spitzer} observations). 

[\OI] and OH lines also contribute to the gas cooling. 
In particular, both the absolute OH luminosities and the observed
 [\OI]63\,$\mu$m/H$_2$ 0-0 $S$(1)\,$\gg$1 intensity ratio
are too bright for non-dissociative  $C$-shock models 
\citep{Kau96,Flo10}. As noted before,
a significant fraction of the [\OI] and OH line emission likely contributes
 to the cooling behind a fast, dissociative $J$-type shock.

\subsection{Shock and UV-driven Chemistry}

The plethora of atomic fine-structure lines   and to a lesser extent,
the relatively high OH/H$_2$O$\lesssim$0.5 abundance ratio inferred in the \textit{warm}+\textit{hot} component, 
confirms the presence of strong dissociative shocks in the inner envelope. 
In addition, owing to efficient H$_2$O photodissociation, an enhanced UV radiation field illuminating the shocked gas
can produce a high OH/H$_2$O abundance ratios  
\citep[\textit{cf.}][in the Orion Bar~PDR]{Goi11}.
The UV radiation field near low-mass protostars (roughly a diluted $<$10,000\,K blackbody)
 is thought to be dominated by Ly\,$\alpha$ photons (10.2\,eV) arising
from accreting material onto the star \citep{Ber03} and from fast dissociative $J$-shocks \citep{Hol79}
such as those
inferred in Serpens\,SMM1 (Section~5.1.1). 
Ly\,$\alpha$  radiation can dissociate H$_2$O (producing enhanced OH and O) but can not dissociate
CO ($\sim$11.1\,eV) nor ionize  sulfur (10.4\,eV) or carbon (11.3\,eV) atoms. 
However,  if significant C$^+$ and H$_2$ are present,
the ion-neutral drift in  $C$-type shocks would significantly enhance the CH$^+$ 
formation rate compared to $J$-type shocks 
\citep[see][for the detection of CH$^+$ $J$=1-0 broad outflow wings in the
shock associated with the DR\,21 massive star forming region]{Fal10}.
The lack of  CH$^+$ $J$=3-2 emission in Serpens\,SMM1 provides an upper limit for the  CH$^+$ column 
density in the shocked gas  (CH$^+$/H$_2$O$<$0.01) and will help to constrain  
UV-irradiated shocks models (Kaufman et al. in prep.; Lesaffre et al. in prep). 
The CH$^+$ $J$=3-2 line is detected in PDRs illuminated by massive OB stars  
where C$^+$ is the dominant carbon species \citep[][Nagy et al. in prep.]{Goi11} and, as in DR\,21, the UV radiation field contains photons
with energies  higher than  Ly\,$\alpha$ (up to 13.6\,eV).
Excited lines from reactive ions such as CO$^+$ (that forms by reaction of C$^+$ with OH) and are
related with the presence of strong UV radiation fields
in the shocked gas,  are not detected in the PACS spectrum  despite previous tentative
 assignations of several high-energy far-IR lines
\citep[see][for IRAS~16293-2422]{Cec97}

The spatial distribution of the [\CII]158\,$\mu$m line emission detected towards Serpens\,SMM1 
(Figure~4)
is similar to that of other species in the outflow (with brighter emission in the outflow position
than towards the protostar itself). The  detection of faint C$^+$ emission indicates the presence of a  relatively weak UV field 
(but able to ionize C atoms and dissociate CO) along the outflow.
Besides, the increase of the 
HCO$^+$/HCN abundance ratio in 
the \textit{warm} temperature component
(HCN is easily photodissociated while HCO$^+$ is abundant
in the dense gas directly exposed to strong UV radiation) is another signature of UV photons.
Indeed, the mere detection of [\CI]370, 609\,$\mu$m lines  at the poor spectral resolution
of the SPIRE-FTS towards the protostar shows that [\CI] lines are significantly brighter than towards
HH46 \citep{vK09} or NGC1333\,IRAS~4A/4B \citep{Yil12}.
The presence of weak C$^+$ line emission and of Ly\,$\alpha$ radiation in protostars like Serpens\,SMM1 (although difficult to detect)  suggests 
\textit{that UV-irradiated shocks are a common phenomenon in YSOs}. 

X-ray emission is also expected in low-mass YSOs \citep{Sta06,Sta07,Fei10} and 
they produce \textit{internally} generated
UV radiation fields after collisions of energetic photoelectrons with H and H$_2$ 
\citep{Mal96,Hol09}.
Although X-ray detections  with \textit{Chandra}  have been reported towards the 
nearby Serpens\,SMM5, SMM6 and S68Nb protostars \citep{Win07},
the strength of any X-ray emission from Serpens\,SMM1 is unknown, possibly because of the high column 
density to the embedded source.

High sensitivity and velocity-resolved observations of CH$^+$, CO$^+$, SO$^+$ and other reactive ions
related with the presence of ionized atoms  (\textit{e.g.,}~C$^+$ and S$^+$)
will help us to characterize these UV-irradiated shocks in more detail.

\subsection{Comparison with Other Low-Mass Protostars}

Comparative spectroscopy of protostars in different stages of evolution allows us to identify the common and 
the more peculiar processes associated with the first stages of star formation.
Comparing the far-IR spectrum of the NGC~1333 IRAS~4B outflow to that of the Serpens\,SMM1 protostar, both show similar high 
H$_2$O luminosities ($L$(H$_2$O)$\sim$0.03\,L$_{\odot}$). However, Serpens\,SMM1 
shows a factor $\sim$50 stronger [\OI] luminosity and a lower $L$(H$_2$O)/$L$(CO) ratio.
The weak [\OI] emission and high H$_2$O luminosity 
in IRAS~4B outflow ($L$(H$_2$O)/$L$(CO)$\simeq$1) has been interpreted as non-dissociative $C$-shocks
shielded from UV radiation \citep{Her12}.
Indeed, C$^+$ is not detected in IRAS~4B outflow whereas it is  detected in Serpens\,SMM1 protostar
and outflow (see Figure~4).
On the other hand, the strong [\OI] and OH emission towards the Serpens\,SMM1 protostar itself 
 is more similar to that of  Class~I source HH46 \citep{vK10} where a $L$(OH)/$L$(H$_2$O)$\simeq$0.5 
luminosity ratio has been inferred (vs. $\sim$0.4 in SMM1). 
One possibility is that a high-velocity jet impinging on the dense 
 inner envelope produces dissociative $J$-shocks \citep{vK10} 
and enhanced [\OI] and OH emission
compared to non-dissociative $C$-shocks. 
Although they could not distinguish the dominant scenario, either
$J$-shocks or UV-irradiated $C$-shocks have been also proposed to explain the OH emission seen 
in the high-mass YSO W3~IRS~5 \citep{Wam11}.

The detection of
[\NeII], [\FeII], [\SiII] and [\SI] fine-structure lines towards Serpens\,SMM1 reinforces the scenario
of both fast and slow shocks as well as UV radiation near the protostar, however, 
it is difficult to extract the exact
geometry of the different shock components in the circumstellar environment (\textit{e.g.,} Ne$^+$ versus H$_2$O
line emitting regions).
Note that the presence of an embedded jet in the Class~0 source L1448 was reported from the detection
of [\FeII] and [\SiII] lines 
by Dionatos et al. (2009), and the same lines have been detected towards the nearby YSOs Serpens\,SMM3 and SMM4 \citep{Lah10}. 
However, they did not detect the [\NeII]12.8\,$\mu$m line that requires fast shock velocities
\citep{Lah07,Bal12}
or  X-ray radiation \citep[see][for Class~II disk sources]{Gue10}.

Compared to HH46,  CO lines with $J_u$$>$30 and higher excitation H$_2$O lines  are detected in Serpens\,SMM1
(not necessarily due to different excitation conditions but maybe just because  lines are much brighter in SMM1). 
We have proposed that this \textit{hot} CO and H$_2$O emission 
arises from low velocity, non-dissociative shocks  
in the inner walls of the outflow cavity  
compressing the gas to very high thermal pressures ($P/k$$=$ $n$\,$T_{\rm K}$$\simeq$10$^{9-10}$~K\,cm$^{-3}$).
Based on the large gas compression factors  and H$_2$O line profiles seen toward
several shock spots in bipolar outflows  (far from the protostellar sources)
Santangelo et al. (2012) and Tafalla et al. (in prep.) also
conclude that current low velocity $J$-shocks models explain their observations better than stationary
$C$-shocks models.
New shock models with a more detailed geometrical description and including the effects of UV radiation
are clearly needed to determine the exact nature of the shocks inferred 
in the dense circumstellar gas  near protostars.

The \textit{warm} CO line emission (15$\lesssim$$J_u$$\lesssim$25) observed in Serpens SMM1 is  a common feature in
the far-IR spectrum of most protostars even if they are in different stages of evolution.
This was  previously observed by ISO \citep[\textit{e.g.,}][]{Gia01} and now by
 \textit{Herschel} (\textit{e.g.,} Karska et al. submitted.; Green et al. in prep., Manoj et al., in prep.). 
This   CO  emission \textit{ubiquity} necessarily means that  a broad combination of shock velocities and densities
can produce a similar \textit{warm} CO spectrum in different low- and high-mass YSOs.

The $L$(OH)/$L$(H$_2$O), $L$(CO)/$L$([\OI]) and $L$(H$_2$O)/$L$([\OI])  luminosity ratios
in Serpens\,SMM1 have intermediate
values between those in NGC~1333 IRAS~4B~Class~0 YSO and HH46~~Class~I YSO (with ratios closer to SMM1) suggesting 
that the different luminosity ratios in IRAS~4B outflow  are either a consequence of an earlier stage of evolution, 
or simply because the IRAS 4B outflow is  peculiar, with very high H$_2$O abundances not affected by UV photodissociation.
Indeed, one of the main conclusions of all these studies is that \textit{water vapour lines are a roboust probe 
of shocked gas in protostellar environments}. In Serpens\,SMM1 we infer a  total
H$_2$O$_h$/(CO$_h$+CO$_w$)$\simeq$0.02 abundance ratio
in the \textit{warm}+\textit{hot} components traced by our  far-IR observations. 
Only if the \textit{hot} component was an independent physical component,
the H$_2$O abundances would be much higher (H$_2$O$_h$/CO$_h$$\simeq$0.4).
However, this value is higher than the typical H$_2$O/CO abundance ratios inferred from HIFI observations of the 
ground-state $o$-H$_2$O 1$_{10}$-1$_{01}$ line (at $\sim$557\,GHz) towards a large sample of low--mass protostars \citep{Kri12}.
In these velocity-resolved observations, the H$_2$O/CO abundance ratio increases from $\sim$10$^{-3}$
at low  outflow velocities ($<$5\,km\,s$^{-1}$) to $\sim$0.1 at high outflow velocities ($>$10 \,km\,s$^{-1}$).
Therefore, our inferred abundance ratio of H$_2$O$_h$/(CO$_h$+CO$_w$)$\simeq$0.02 from PACS observations seems a better description of the water
vapour abundance in the protostellar shocked gas. Besides, it is also consistent with recent determinations from velocity-resolved
HIFI observations of
moderately excited  H$_2$O lines ($E_{\rm u}$/$k$$<$215\,K)
\citep[see \textit{e.g.,}][for L1448 outflow]{San12}.

\vspace{-0.2cm}
\section{Summary and Conclusions}

We have presented the first complete far-IR and submm spectrum of a Class~0 protostar (Serpens\,SMM1) taken
with \textit{Herschel} in the $\sim$55-671\,$\mu$m window. 
The data are complemented with mid-IR spectra in the $\sim$10-37\,$\mu$m window 
retrieved  from the \textit{Spitzer} archive and first discussed here. 
These spectroscopic observations span almost 2 decades in wavelength and allow us to unveil the most 
important heating and chemical processes associated
with the first evolutionary  stages of  low-mass protostars. 
In particular, we obtained the following results:\\

$\bullet$~Serpens\,SMM1 shows a very rich far-IR and submm emission spectrum, with 
more than 145 lines, most of them rotationally excited lines of $^{12}$CO (up to $J$=42-41), 
H$_2$O, OH, $^{13}$CO, HCN, HCO$^+$ and weaker emission from [\CII]158 and [\CI]370, 609\,$\mu$m.
[\OI]63\,$\mu$m is the brightest line.
Approximately half of the total far-IR and submm line luminosity is provided by CO rotational emission,
 with important contribution from
H$_2$O ($\sim$22\%), [\OI] ($\sim$12\%) and OH ($\sim$9\%). 

$\bullet$~The mid-IR spectrum shows many fewer lines  (probably due to severe dust extinction and lack of  spectral resolution).
Bright fine structure lines from  Ne$^+$, Fe$^+$,  Si$^+$, and S, 
as well as weak H$_2$ $S$(1) and $S$(2) pure  rotational lines  are detected. %(up to $E_{\rm u}$/$k$$=$1682\,K).  

$\bullet$~The $^{12}$CO rotational diagram suggests the presence of 3 temperature components with
$T$$_{\rm rot}^{hot}$$\simeq$620$\pm$30\,K, $T$$_{\rm rot}^{warm}$$\simeq$340$\pm$40\,K  
and $T$$_{\rm rot}^{cool}$$\simeq$100$\pm$15\,K. 
As predicted by SED models in the literature, 
the detection of H$_2$O, OH, HCO$^+$ and HCN emission lines from very high critical density transitions 
suggests that the density in the inner envelope is high ($n$(H$_2$)$\gtrsim$5$\times$10$^{6}$\,cm$^{-3}$), and thus the CO rotational
temperatures provide a good approximation to the gas temperature ($T$$_{\rm rot}$$\lesssim$$T$$_{\rm k}$).

$\bullet$~A  non-LTE,  multi-component   model allowed us
to approximately quantify the contribution of the different temperature components 
($T$$_{\rm k}^{hot}$$\approx$800\,K, $T$$_{\rm k}^{warm}$$\approx$375\,K and $T_{\rm k}^{cool}$$\approx$150\,K)
and to estimate relative
 chemical abundances.  
The detected mid-IR  H$_2$ lines  provide a lower limit 
to H$_2$ column density of $N_{\rm H_2}$$\geq$10$^{22}$\,cm$^{-2}$.
We derive the following upper limit abundances with respect to H$_2$  in the 
\textit{hot}+\textit{warm} components: 
$x$(CO)$\leq$10$^{-4}$, $x$(H$_2$O)$\leq$0.5$\times$10$^{-5}$ and 
 $x$(OH)$\leq$10$^{-6}$. The inferred water vapour abundance is higher than the  value
typically found in cold interstellar clouds but lower than that expected in hot shocked gas 
completely shielded from UV radiation

$\bullet$~Excited CO, H$_2$O and OH emission lines  arising from high energy levels are detected 
(up to $E_{\rm u}$/$k$$=$4971\,K, $E_{\rm u}$/$k$$=$1036\,K and $E_{\rm u}$/$k$$=$618\,K respectively). 
These species arise in the  \textit{hot}+\textit{warm} gas ($M$$\lesssim$0.03\,$M$$_{\odot}$) that we associate with 
a mixture of shocks. 
The observed \textit{hot} CO, H$_2$O and H$_2$ lines provide a lower limit to the gas temperature
 of  $\sim$700\,K. 
The excited H$_2$O line emission is consistent with detailed model predictions of  
low-velocity non-dissociative 
shocks (with v$_{\rm s}$$\lesssim$20\,km\,s$^{-1}$).

$\bullet$~Fast dissociative (and ionizing) shocks with velocities v$_{\rm s}$$>$60\,km\,s$^{-1}$ and
pre-shock densities $\geq$10$^4$\,cm$^{-3}$ related with the presence of an  embedded  atomic jet 
are needed to explain the observed Ne$^+$, Fe$^+$, Si$^+$ and S fine-structure emission
and also the bright and velocity-shifted [\OI]63\,$\mu$m line emission. 
Dissociative $J$-shocks produced by the jet impacting the ambient material are the most
 probable origin of the bright [\OI] and OH emission and of a significant fraction
of the \textit{warm} CO emission.
In addition,
water vapour photodissociation in UV-irradiated non-dissociative shocks along the outflow cavity walls can
also contribute to the [\OI] and OH emission.\\

Compared to other protostars, the large number of lines detected 
 towards 
Serpens\,SMM1 reveals the great complexity of the protostellar environment. Both fast 
and slow shocks are needed 
to explain  the presence of atomic fine structure lines and high excitation molecular lines.
The spectra also show the signature of UV radiation (C$^+$ and C towards the protostar and outflow).
Therefore, most species (including H$_2$O, OH, O and trace molecules such as CH$^+$) 
arise in shocked gas illuminated by UV (even X-ray) radiation fields. 
Irradiated shocks are likely a very common phenomenon in YSOs.

\begin{acknowledgements}
We acknowledge our WISH internal referees, D. Neufeld and P. Bjerkeli
for very helpful comments on an earlier version of the manuscript.
We also thank the entire WISH team for many useful and vivid discussions in the last years.
We finally thank the anonymous referee and M.~Walmsley for useful comments.
WISH research in Leiden is supported by the Netherlands Research
School for Astronomy (NOVA), by a Spinoza grant and grant 614.001.008
from the Netherlands Organisation for Scientific Research (NWO), 
and by EU-FP7 grant 238258 (LASSIE).
JRC, JC and ME thank the Spanish MINECO for funding support
through grants AYA2009-07304 and CSD2009-00038. 
JRG is supported by a Ram\'on y Cajal research contract from the MINECO
and co-financed by the European Social Fund.

\end{acknowledgements}

\bibliographystyle{aa}
\bibliography{19912}

\begin{thebibliography}{63}
\expandafter\ifx\csname natexlab\endcsname\relax\def\natexlab#1{#1}\fi

\bibitem[{{Bachiller} \& {Tafalla}(1999)}]{Bac99}
{Bachiller}, R. \& {Tafalla}, M. 1999, in NATO ASIC Proc. 540: The Origin of
  Stars and Planetary Systems, ed. C.~J. {Lada} \& N.~D. {Kylafis}, 227

\bibitem[{{Baldovin-Saavedra} {et~al.}(2012){Baldovin-Saavedra}, {Audard},
  {Carmona}, {G{\"u}del}, {Briggs}, {Rebull}, {Skinner}, \& {Ercolano}}]{Bal12}
{Baldovin-Saavedra}, C., {Audard}, M., {Carmona}, A., {et~al.} 2012, \aap, 543,
  A30

\bibitem[{{Bergin} {et~al.}(2003){Bergin}, {Calvet}, {D'Alessio}, \&
  {Herczeg}}]{Ber03}
{Bergin}, E., {Calvet}, N., {D'Alessio}, P., \& {Herczeg}, G.~J. 2003, \apjl,
  591, L159

\bibitem[{{Bontemps} {et~al.}(1996){Bontemps}, {Andre}, {Terebey}, \&
  {Cabrit}}]{Bon96}
{Bontemps}, S., {Andre}, P., {Terebey}, S., \& {Cabrit}, S. 1996, \aap, 311,
  858

\bibitem[{{Caselli} {et~al.}(2010){Caselli}, {Keto}, {Pagani}, {Aikawa},
  {Y{\i}ld{\i}z}, {van der Tak}, {Tafalla}, {Bergin}, {Nisini}, {Codella}, {van
  Dishoeck}, {Bachiller}, {Baudry}, {Benedettini}, {Benz}, {Bjerkeli}, {Blake},
  {Bontemps}, {Braine}, {Bruderer}, {Cernicharo}, {Daniel}, {di Giorgio},
  {Dominik}, {Doty}, {Encrenaz}, {Fich}, {Fuente}, {Gaier}, {Giannini},
  {Goicoechea}, {de Graauw}, {Helmich}, {Herczeg}, {Herpin}, {Hogerheijde},
  {Jackson}, {Jacq}, {Javadi}, {Johnstone}, {J{\o}rgensen}, {Kester},
  {Kristensen}, {Laauwen}, {Larsson}, {Lis}, {Liseau}, {Luinge}, {Marseille},
  {McCoey}, {Megej}, {Melnick}, {Neufeld}, {Olberg}, {Parise}, {Pearson},
  {Plume}, {Risacher}, {Santiago-Garc{\'{\i}}a}, {Saraceno}, {Shipman},
  {Siegel}, {van Kempen}, {Visser}, {Wampfler}, \& {Wyrowski}}]{Cas10}
{Caselli}, P., {Keto}, E., {Pagani}, L., {et~al.} 2010, \aap, 521, L29

\bibitem[{{Ceccarelli} {et~al.}(1997){Ceccarelli}, {Caux}, {Wolfire},
  {Rudolph}, {Nisini}, {Saraceno}, \& {White}}]{Cec97}
{Ceccarelli}, C., {Caux}, E., {Wolfire}, M., {et~al.} 1997, in ESA Special
  Publication, Vol. 419, The first ISO workshop on Analytical Spectroscopy, ed.
  A.~M. {Heras}, K.~{Leech}, N.~R. {Trams}, \& M.~{Perry}, 43

\bibitem[{{Cernicharo}(2012)}]{Cer12}
{Cernicharo}, J. 2012, in European Astronomical Society Publications Series,
  Vol. 2012, Proceedings of the European Conference on Laboratory Astrophysics,
  ed. C.~{Stehl\'e}, C.~{Joblin}, \& L.~{d'Hendecourt}, 4

\bibitem[{{Cernicharo} {et~al.}(2006){Cernicharo}, {Goicoechea}, {Daniel},
  {Lerate}, {Barlow}, {Swinyard}, {van Dishoeck}, {Lim}, {Viti}, \&
  {Yates}}]{Cer06}
{Cernicharo}, J., {Goicoechea}, J.~R., {Daniel}, F., {et~al.} 2006, \apjl, 649,
  L33

\bibitem[{{Chiar} {et~al.}(2007){Chiar}, {Ennico}, {Pendleton}, {Boogert},
  {Greene}, {Knez}, {Lada}, {Roellig}, {Tielens}, {Werner}, \&
  {Whittet}}]{Chi07}
{Chiar}, J.~E., {Ennico}, K., {Pendleton}, Y.~J., {et~al.} 2007, \apjl, 666,
  L73

\bibitem[{{Davis} {et~al.}(1999){Davis}, {Matthews}, {Ray}, {Dent}, \&
  {Richer}}]{Dav99}
{Davis}, C.~J., {Matthews}, H.~E., {Ray}, T.~P., {Dent}, W.~R.~F., \& {Richer},
  J.~S. 1999, \mnras, 309, 141

\bibitem[{{Dionatos} {et~al.}(2010){Dionatos}, {Nisini}, {Codella}, \&
  {Giannini}}]{Dio10}
{Dionatos}, O., {Nisini}, B., {Codella}, C., \& {Giannini}, T. 2010, \aap, 523,
  A29

\bibitem[{{Dionatos} {et~al.}(2009){Dionatos}, {Nisini}, {Garcia Lopez},
  {Giannini}, {Davis}, {Smith}, {Ray}, \& {DeLuca}}]{Dio09}
{Dionatos}, O., {Nisini}, B., {Garcia Lopez}, R., {et~al.} 2009, \apj, 692, 1

\bibitem[{{Draine} \& {McKee}(1993)}]{Dra93}
{Draine}, B.~T. \& {McKee}, C.~F. 1993, \araa, 31, 373

\bibitem[{{Dzib} {et~al.}(2010){Dzib}, {Loinard}, {Mioduszewski}, {Boden},
  {Rodr{\'{\i}}guez}, \& {Torres}}]{Dzi10}
{Dzib}, S., {Loinard}, L., {Mioduszewski}, A.~J., {et~al.} 2010, \apj, 718, 610

\bibitem[{{Eiroa} {et~al.}(2008){Eiroa}, {Djupvik}, \& {Casali}}]{Eir08}
{Eiroa}, C., {Djupvik}, A.~A., \& {Casali}, M.~M. 2008, {The Serpens Molecular
  Cloud}, ed. B.~{Reipurth}, 693

\bibitem[{{Enoch} {et~al.}(2009){Enoch}, {Corder}, {Dunham}, \&
  {Duch{\^e}ne}}]{Eno09}
{Enoch}, M.~L., {Corder}, S., {Dunham}, M.~M., \& {Duch{\^e}ne}, G. 2009, \apj,
  707, 103

\bibitem[{{Evans} {et~al.}(2009){Evans}, {Dunham}, {J{\o}rgensen}, {Enoch},
  {Mer{\'{\i}}n}, {van Dishoeck}, {Alcal{\'a}}, {Myers}, {Stapelfeldt},
  {Huard}, {Allen}, {Harvey}, {van Kempen}, {Blake}, {Koerner}, {Mundy},
  {Padgett}, \& {Sargent}}]{Eva09}
{Evans}, II, N.~J., {Dunham}, M.~M., {J{\o}rgensen}, J.~K., {et~al.} 2009,
  \apjs, 181, 321

\bibitem[{{Falgarone} {et~al.}(2010){Falgarone}, {Ossenkopf}, {Gerin},
  {Lesaffre}, {Godard}, {Pearson}, {Cabrit}, {Joblin}, {Benz}, {Boulanger},
  {Fuente}, {G{\"u}sten}, {Harris}, {Klein}, {Kramer}, {Lord}, {Martin},
  {Martin-Pintado}, {Neufeld}, {Phillips}, {R{\"o}llig}, {Simon}, {Stutzki},
  {van der Tak}, {Teyssier}, {Yorke}, {Erickson}, {Fich}, {Jellema}, {Marston},
  {Risacher}, {Salez}, \& {Schm{\"u}lling}}]{Fal10}
{Falgarone}, E., {Ossenkopf}, V., {Gerin}, M., {et~al.} 2010, \aap, 518, L118

\bibitem[{{Feigelson}(2010)}]{Fei10}
{Feigelson}, E.~D. 2010, Proceedings of the National Academy of Science, 107,
  7153

\bibitem[{{Flower} \& {Pineau Des For{\^e}ts}(2010)}]{Flo10}
{Flower}, D.~R. \& {Pineau Des For{\^e}ts}, G. 2010, \mnras, 406, 1745

\bibitem[{{Giannini} {et~al.}(2001){Giannini}, {Nisini}, \&
  {Lorenzetti}}]{Gia01}
{Giannini}, T., {Nisini}, B., \& {Lorenzetti}, D. 2001, \apj, 555, 40

\bibitem[{{Goicoechea} \& {Cernicharo}(2002)}]{Goi02}
{Goicoechea}, J.~R. \& {Cernicharo}, J. 2002, \apjl, 576, L77

\bibitem[{{Goicoechea} {et~al.}(2006){Goicoechea}, {Cernicharo}, {Lerate},
  {Daniel}, {Barlow}, {Swinyard}, {Lim}, {Viti}, \& {Yates}}]{Goi06}
{Goicoechea}, J.~R., {Cernicharo}, J., {Lerate}, M.~R., {et~al.} 2006, \apjl,
  641, L49

\bibitem[{{Goicoechea} {et~al.}(2011){Goicoechea}, {Joblin}, {Contursi},
  {Bern{\'e}}, {Cernicharo}, {Gerin}, {Le Bourlot}, {Bergin}, {Bell}, \&
  {R{\"o}llig}}]{Goi11}
{Goicoechea}, J.~R., {Joblin}, C., {Contursi}, A., {et~al.} 2011, \aap, 530,
  L16

\bibitem[{{Griffin} {et~al.}(2010){Griffin}, {Abergel}, {Abreu}, {Ade},
  {Andr{\'e}}, {Augueres}, {Babbedge}, {Bae}, {Baillie}, {Baluteau}, {Barlow},
  {Bendo}, {Benielli}, {Bock}, {Bonhomme}, {Brisbin}, {Brockley-Blatt},
  {Caldwell}, {Cara}, {Castro-Rodriguez}, {Cerulli}, {Chanial}, {Chen},
  {Clark}, {Clements}, {Clerc}, {Coker}, {Communal}, {Conversi}, {Cox},
  {Crumb}, {Cunningham}, {Daly}, {Davis}, {de Antoni}, {Delderfield}, {Devin},
  {di Giorgio}, {Didschuns}, {Dohlen}, {Donati}, {Dowell}, {Dowell}, {Duband},
  {Dumaye}, {Emery}, {Ferlet}, {Ferrand}, {Fontignie}, {Fox}, {Franceschini},
  {Frerking}, {Fulton}, {Garcia}, {Gastaud}, {Gear}, {Glenn}, {Goizel},
  {Griffin}, {Grundy}, {Guest}, {Guillemet}, {Hargrave}, {Harwit}, {Hastings},
  {Hatziminaoglou}, {Herman}, {Hinde}, {Hristov}, {Huang}, {Imhof}, {Isaak},
  {Israelsson}, {Ivison}, {Jennings}, {Kiernan}, {King}, {Lange}, {Latter},
  {Laurent}, {Laurent}, {Leeks}, {Lellouch}, {Levenson}, {Li}, {Li},
  {Lilienthal}, {Lim}, {Liu}, {Lu}, {Madden}, {Mainetti}, {Marliani}, {McKay},
  {Mercier}, {Molinari}, {Morris}, {Moseley}, {Mulder}, {Mur}, {Naylor},
  {Nguyen}, {O'Halloran}, {Oliver}, {Olofsson}, {Olofsson}, {Orfei}, {Page},
  {Pain}, {Panuzzo}, {Papageorgiou}, {Parks}, {Parr-Burman}, {Pearce},
  {Pearson}, {P{\'e}rez-Fournon}, {Pinsard}, {Pisano}, {Podosek}, {Pohlen},
  {Polehampton}, {Pouliquen}, {Rigopoulou}, {Rizzo}, {Roseboom}, {Roussel},
  {Rowan-Robinson}, {Rownd}, {Saraceno}, {Sauvage}, {Savage}, {Savini},
  {Sawyer}, {Scharmberg}, {Schmitt}, {Schneider}, {Schulz}, {Schwartz},
  {Shafer}, {Shupe}, {Sibthorpe}, {Sidher}, {Smith}, {Smith}, {Smith},
  {Spencer}, {Stobie}, {Sudiwala}, {Sukhatme}, {Surace}, {Stevens}, {Swinyard},
  {Trichas}, {Tourette}, {Triou}, {Tseng}, {Tucker}, {Turner}, {Vaccari},
  {Valtchanov}, {Vigroux}, {Virique}, {Voellmer}, {Walker}, {Ward}, {Waskett},
  {Weilert}, {Wesson}, {White}, {Whitehouse}, {Wilson}, {Winter}, {Woodcraft},
  {Wright}, {Xu}, {Zavagno}, {Zemcov}, {Zhang}, \& {Zonca}}]{Grif10}
{Griffin}, M.~J., {Abergel}, A., {Abreu}, A., {et~al.} 2010, \aap, 518, L3

\bibitem[{{G{\"u}del} {et~al.}(2010){G{\"u}del}, {Lahuis}, {Briggs}, {Carr},
  {Glassgold}, {Henning}, {Najita}, {van Boekel}, \& {van Dishoeck}}]{Gue10}
{G{\"u}del}, M., {Lahuis}, F., {Briggs}, K.~R., {et~al.} 2010, \aap, 519, A113

\bibitem[{{Herczeg} {et~al.}(2012){Herczeg}, {Karska}, {Bruderer},
  {Kristensen}, {van Dishoeck}, {J{\o}rgensen}, {Visser}, {Wampfler}, {Bergin},
  {Y{\i}ld{\i}z}, {Pontoppidan}, \& {Gracia-Carpio}}]{Her12}
{Herczeg}, G.~J., {Karska}, A., {Bruderer}, S., {et~al.} 2012, \aap, 540, A84

\bibitem[{{Hogerheijde} {et~al.}(1999){Hogerheijde}, {van Dishoeck},
  {Salverda}, \& {Blake}}]{Hog99}
{Hogerheijde}, M.~R., {van Dishoeck}, E.~F., {Salverda}, J.~M., \& {Blake},
  G.~A. 1999, \apj, 513, 350

\bibitem[{{Hollenbach} \& {Gorti}(2009)}]{Hol09}
{Hollenbach}, D. \& {Gorti}, U. 2009, \apj, 703, 1203

\bibitem[{{Hollenbach} \& {McKee}(1979)}]{Hol79}
{Hollenbach}, D. \& {McKee}, C.~F. 1979, \apjs, 41, 555

\bibitem[{{Hollenbach} \& {McKee}(1989)}]{Hol89}
{Hollenbach}, D. \& {McKee}, C.~F. 1989, \apj, 342, 306

\bibitem[{{Houck} {et~al.}(2004){Houck}, {Roellig}, {van Cleve}, {Forrest},
  {Herter}, {Lawrence}, {Matthews}, {Reitsema}, {Soifer}, {Watson}, {Weedman},
  {Huisjen}, {Troeltzsch}, {Barry}, {Bernard-Salas}, {Blacken}, {Brandl},
  {Charmandaris}, {Devost}, {Gull}, {Hall}, {Henderson}, {Higdon}, {Pirger},
  {Schoenwald}, {Sloan}, {Uchida}, {Appleton}, {Armus}, {Burgdorf},
  {Fajardo-Acosta}, {Grillmair}, {Ingalls}, {Morris}, \& {Teplitz}}]{Hou04}
{Houck}, J.~R., {Roellig}, T.~L., {van Cleve}, J., {et~al.} 2004, \apjs, 154,
  18

\bibitem[{{Kaufman} \& {Neufeld}(1996)}]{Kau96}
{Kaufman}, M.~J. \& {Neufeld}, D.~A. 1996, \apj, 456, 611

\bibitem[{{Kristensen} {et~al.}(2012){Kristensen}, {van Dishoeck}, {Bergin},
  {Visser}, {Y{\i}ld{\i}z}, {San Jose-Garcia}, {J{\o}rgensen}, {Herczeg},
  {Johnstone}, {Wampfler}, {Benz}, {Bruderer}, {Cabrit}, {Caselli}, {Doty},
  {Harsono}, {Herpin}, {Hogerheijde}, {Karska}, {van Kempen}, {Liseau},
  {Nisini}, {Tafalla}, {van der Tak}, \& {Wyrowski}}]{Kri12}
{Kristensen}, L.~E., {van Dishoeck}, E.~F., {Bergin}, E.~A., {et~al.} 2012,
  \aap, 542, A8

\bibitem[{{Lahuis} {et~al.}(2007){Lahuis}, {van Dishoeck}, {Blake}, {Evans},
  {Kessler-Silacci}, \& {Pontoppidan}}]{Lah07}
{Lahuis}, F., {van Dishoeck}, E.~F., {Blake}, G.~A., {et~al.} 2007, \apj, 665,
  492

\bibitem[{{Lahuis} {et~al.}(2010){Lahuis}, {van Dishoeck}, {J{\o}rgensen},
  {Blake}, \& {Evans}}]{Lah10}
{Lahuis}, F., {van Dishoeck}, E.~F., {J{\o}rgensen}, J.~K., {Blake}, G.~A., \&
  {Evans}, N.~J. 2010, \aap, 519, A3

\bibitem[{{Langer} \& {Penzias}(1990)}]{Lan90}
{Langer}, W.~D. \& {Penzias}, A.~A. 1990, \apj, 357, 477

\bibitem[{{Laor} \& {Draine}(1993)}]{Lao93}
{Laor}, A. \& {Draine}, B.~T. 1993, \apj, 402, 441

\bibitem[{{Larsson} {et~al.}(2002){Larsson}, {Liseau}, \&
  {Men'shchikov}}]{Lar02}
{Larsson}, B., {Liseau}, R., \& {Men'shchikov}, A.~B. 2002, \aap, 386, 1055

\bibitem[{{Larsson} {et~al.}(2000){Larsson}, {Liseau}, {Men'shchikov},
  {Olofsson}, {Caux}, {Ceccarelli}, {Lorenzetti}, {Molinari}, {Nisini},
  {Nordh}, {Saraceno}, {Sibille}, {Spinoglio}, \& {White}}]{Lar00}
{Larsson}, B., {Liseau}, R., {Men'shchikov}, A.~B., {et~al.} 2000, \aap, 363,
  253

\bibitem[{{Maloney} {et~al.}(1996){Maloney}, {Hollenbach}, \&
  {Tielens}}]{Mal96}
{Maloney}, P.~R., {Hollenbach}, D.~J., \& {Tielens}, A.~G.~G.~M. 1996, \apj,
  466, 561

\bibitem[{{Melnick} {et~al.}(2008){Melnick}, {Tolls}, {Neufeld}, {Yuan},
  {Sonnentrucker}, {Watson}, {Bergin}, \& {Kaufman}}]{Mel08}
{Melnick}, G.~J., {Tolls}, V., {Neufeld}, D.~A., {et~al.} 2008, \apj, 683, 876

\bibitem[{{Neufeld} \& {Dalgarno}(1989)}]{Neu89}
{Neufeld}, D.~A. \& {Dalgarno}, A. 1989, \apj, 340, 869

\bibitem[{{Neufeld} {et~al.}(2006){Neufeld}, {Melnick}, {Sonnentrucker},
  {Bergin}, {Green}, {Kim}, {Watson}, {Forrest}, \& {Pipher}}]{Neu06}
{Neufeld}, D.~A., {Melnick}, G.~J., {Sonnentrucker}, P., {et~al.} 2006, \apj,
  649, 816

\bibitem[{{Nisini} {et~al.}(2005){Nisini}, {Bacciotti}, {Giannini}, {Massi},
  {Eisl{\"o}ffel}, {Podio}, \& {Ray}}]{Nis05}
{Nisini}, B., {Bacciotti}, F., {Giannini}, T., {et~al.} 2005, \aap, 441, 159

\bibitem[{{Offer} \& {van Dishoeck}(1992)}]{Off92}
{Offer}, A.~R. \& {van Dishoeck}, E.~F. 1992, \mnras, 257, 377

\bibitem[{{Pilbratt} {et~al.}(2010){Pilbratt}, {Riedinger}, {Passvogel},
  {Crone}, {Doyle}, {Gageur}, {Heras}, {Jewell}, {Metcalfe}, {Ott}, \&
  {Schmidt}}]{Pil10}
{Pilbratt}, G.~L., {Riedinger}, J.~R., {Passvogel}, T., {et~al.} 2010, \aap,
  518, L1

\bibitem[{{Poglitsch} {et~al.}(2010){Poglitsch}, {Waelkens}, {Geis},
  {Feuchtgruber}, {Vandenbussche}, {Rodriguez}, {Krause}, {Renotte}, {van
  Hoof}, {Saraceno}, {Cepa}, {Kerschbaum}, {Agn{\`e}se}, {Ali}, {Altieri},
  {Andreani}, {Augueres}, {Balog}, {Barl}, {Bauer}, {Belbachir}, {Benedettini},
  {Billot}, {Boulade}, {Bischof}, {Blommaert}, {Callut}, {Cara}, {Cerulli},
  {Cesarsky}, {Contursi}, {Creten}, {De Meester}, {Doublier}, {Doumayrou},
  {Duband}, {Exter}, {Genzel}, {Gillis}, {Gr{\"o}zinger}, {Henning},
  {Herreros}, {Huygen}, {Inguscio}, {Jakob}, {Jamar}, {Jean}, {de Jong},
  {Katterloher}, {Kiss}, {Klaas}, {Lemke}, {Lutz}, {Madden}, {Marquet},
  {Martignac}, {Mazy}, {Merken}, {Montfort}, {Morbidelli}, {M{\"u}ller},
  {Nielbock}, {Okumura}, {Orfei}, {Ottensamer}, {Pezzuto}, {Popesso},
  {Putzeys}, {Regibo}, {Reveret}, {Royer}, {Sauvage}, {Schreiber}, {Stegmaier},
  {Schmitt}, {Schubert}, {Sturm}, {Thiel}, {Tofani}, {Vavrek}, {Wetzstein},
  {Wieprecht}, \& {Wiezorrek}}]{Pog10}
{Poglitsch}, A., {Waelkens}, C., {Geis}, N., {et~al.} 2010, \aap, 518, L2

\bibitem[{{Rodr{\'{\i}}guez} {et~al.}(2005){Rodr{\'{\i}}guez}, {Loinard},
  {D'Alessio}, {Wilner}, \& {Ho}}]{Rod05}
{Rodr{\'{\i}}guez}, L.~F., {Loinard}, L., {D'Alessio}, P., {Wilner}, D.~J., \&
  {Ho}, P.~T.~P. 2005, \apjl, 621, L133

\bibitem[{{Santangelo} {et~al.}(2012){Santangelo}, {Nisini}, {Giannini},
  {Antoniucci}, {Vasta}, {Codella}, {Lorenzani}, {Tafalla}, {Liseau}, {van
  Dishoeck}, \& {Kristensen}}]{San12}
{Santangelo}, G., {Nisini}, B., {Giannini}, T., {et~al.} 2012, \aap, 538, A45

\bibitem[{{Smith} {et~al.}(2007){Smith}, {Armus}, {Dale}, {Roussel}, {Sheth},
  {Buckalew}, {Jarrett}, {Helou}, \& {Kennicutt}}]{Smi07}
{Smith}, J.~D.~T., {Armus}, L., {Dale}, D.~A., {et~al.} 2007, \pasp, 119, 1133

\bibitem[{{St{\"a}uber} {et~al.}(2007){St{\"a}uber}, {Benz}, {J{\o}rgensen},
  {van Dishoeck}, {Doty}, \& {van der Tak}}]{Sta07}
{St{\"a}uber}, P., {Benz}, A.~O., {J{\o}rgensen}, J.~K., {et~al.} 2007, \aap,
  466, 977

\bibitem[{{St{\"a}uber} {et~al.}(2006){St{\"a}uber}, {J{\o}rgensen}, {van
  Dishoeck}, {Doty}, \& {Benz}}]{Sta06}
{St{\"a}uber}, P., {J{\o}rgensen}, J.~K., {van Dishoeck}, E.~F., {Doty}, S.~D.,
  \& {Benz}, A.~O. 2006, \aap, 453, 555

\bibitem[{{van Dishoeck} {et~al.}(2011){van Dishoeck}, {Kristensen}, {Benz},
  {Bergin}, {Caselli}, {Cernicharo}, {Herpin}, {Hogerheijde}, {Johnstone},
  {Liseau}, {Nisini}, {Shipman}, {Tafalla}, {van der Tak}, {Wyrowski},
  {Aikawa}, {Bachiller}, {Baudry}, {Benedettini}, {Bjerkeli}, {Blake},
  {Bontemps}, {Braine}, {Brinch}, {Bruderer}, {Chavarr{\'{\i}}a}, {Codella},
  {Daniel}, {de Graauw}, {Deul}, {di Giorgio}, {Dominik}, {Doty}, {Dubernet},
  {Encrenaz}, {Feuchtgruber}, {Fich}, {Frieswijk}, {Fuente}, {Giannini},
  {Goicoechea}, {Helmich}, {Herczeg}, {Jacq}, {J{\o}rgensen}, {Karska},
  {Kaufman}, {Keto}, {Larsson}, {Lefloch}, {Lis}, {Marseille}, {McCoey},
  {Melnick}, {Neufeld}, {Olberg}, {Pagani}, {Pani{\'c}}, {Parise}, {Pearson},
  {Plume}, {Risacher}, {Salter}, {Santiago-Garc{\'{\i}}a}, {Saraceno},
  {St{\"a}uber}, {van Kempen}, {Visser}, {Viti}, {Walmsley}, {Wampfler}, \&
  {Y{\i}ld{\i}z}}]{vD11}
{van Dishoeck}, E.~F., {Kristensen}, L.~E., {Benz}, A.~O., {et~al.} 2011,
  \pasp, 123, 138

\bibitem[{{van Kempen} {et~al.}(2010){van Kempen}, {Kristensen}, {Herczeg},
  {Visser}, {van Dishoeck}, {Wampfler}, {Bruderer}, {Benz}, {Doty}, {Brinch},
  {Hogerheijde}, {J{\o}rgensen}, {Tafalla}, {Neufeld}, {Bachiller}, {Baudry},
  {Benedettini}, {Bergin}, {Bjerkeli}, {Blake}, {Bontemps}, {Braine},
  {Caselli}, {Cernicharo}, {Codella}, {Daniel}, {di Giorgio}, {Dominik},
  {Encrenaz}, {Fich}, {Fuente}, {Giannini}, {Goicoechea}, {de Graauw},
  {Helmich}, {Herpin}, {Jacq}, {Johnstone}, {Kaufman}, {Larsson}, {Lis},
  {Liseau}, {Marseille}, {McCoey}, {Melnick}, {Nisini}, {Olberg}, {Parise},
  {Pearson}, {Plume}, {Risacher}, {Santiago-Garc{\'{\i}}a}, {Saraceno},
  {Shipman}, {van der Tak}, {Wyrowski}, {Y{\i}ld{\i}z}, {Ciechanowicz},
  {Dubbeldam}, {Glenz}, {Huisman}, {Lin}, {Morris}, {Murphy}, \&
  {Trappe}}]{vK10}
{van Kempen}, T.~A., {Kristensen}, L.~E., {Herczeg}, G.~J., {et~al.} 2010,
  \aap, 518, L121

\bibitem[{{van Kempen} {et~al.}(2009){van Kempen}, {van Dishoeck},
  {Hogerheijde}, \& {G{\"u}sten}}]{vK09}
{van Kempen}, T.~A., {van Dishoeck}, E.~F., {Hogerheijde}, M.~R., \&
  {G{\"u}sten}, R. 2009, \aap, 508, 259

\bibitem[{{Visser} {et~al.}(2012){Visser}, {Kristensen}, {Bruderer}, {van
  Dishoeck}, {Herczeg}, {Brinch}, {Doty}, {Harsono}, \& {Wolfire}}]{Vis12}
{Visser}, R., {Kristensen}, L.~E., {Bruderer}, S., {et~al.} 2012, \aap, 537,
  A55

\bibitem[{{Walmsley} {et~al.}(2005){Walmsley}, {Pineau des For{\^e}ts}, \&
  {Flower}}]{Wal05}
{Walmsley}, M., {Pineau des For{\^e}ts}, G., \& {Flower}, D. 2005, in IAU
  Symposium, Vol. 231, Astrochemistry: Recent Successes and Current Challenges,
  ed. D.~C. {Lis}, G.~A. {Blake}, \& E.~{Herbst}, 135--140

\bibitem[{{Wampfler} {et~al.}(2011){Wampfler}, {Bruderer}, {Kristensen},
  {Chavarr{\'{\i}}a}, {Bergin}, {Benz}, {van Dishoeck}, {Herczeg}, {van der
  Tak}, {Goicoechea}, {Doty}, \& {Herpin}}]{Wam11}
{Wampfler}, S.~F., {Bruderer}, S., {Kristensen}, L.~E., {et~al.} 2011, \aap,
  531, L16

\bibitem[{{Watson} {et~al.}(2007){Watson}, {Bohac}, {Hull}, {Forrest},
  {Furlan}, {Najita}, {Calvet}, {D'Alessio}, {Hartmann}, {Sargent}, {Green},
  {Kim}, \& {Houck}}]{Wat07}
{Watson}, D.~M., {Bohac}, C.~J., {Hull}, C., {et~al.} 2007, \nat, 448, 1026

\bibitem[{{Whittet}(2003)}]{Whi03}
{Whittet}, D.~C.~B., ed. 2003, {Dust in the galactic environment}

\bibitem[{{Winston} {et~al.}(2007){Winston}, {Megeath}, {Wolk}, {Muzerolle},
  {Gutermuth}, {Hora}, {Allen}, {Spitzbart}, {Myers}, \& {Fazio}}]{Win07}
{Winston}, E., {Megeath}, S.~T., {Wolk}, S.~J., {et~al.} 2007, \apj, 669, 493

\bibitem[{{Y{\i}ld{\i}z} {et~al.}(2012){Y{\i}ld{\i}z}, {Kristensen}, {van
  Dishoeck}, {Belloche}, {van Kempen}, {Hogerheijde}, {G{\"u}sten}, \& {van der
  Marel}}]{Yil12}
{Y{\i}ld{\i}z}, U.~A., {Kristensen}, L.~E., {van Dishoeck}, E.~F., {et~al.}
  2012, \aap, 542, A86

\end{thebibliography}

\begin{appendix}

\section{Complementary Figures and Tables}

\clearpage

 \begin{table}[t]
      \caption[]{List of detected lines towards Serpens~SMM1 and line fluxes from  SPIRE and 
                 PACS central spaxels (corrected for extended emission in the case of PACS).
                 The horizontal line near $\sim$300\,$\mu$m shows the transition between the SPIRE SLW and SSW detector arrays
                 (beams between 42$''$ and 17$''$). 
                 The line near $\sim$200\,$\mu$m marks the lines detected with PACS (spaxel size $\sim$9.4$''$). 
                 Lines in the PACS ranges with detector leakage are not tabulated.
                 Flux errors up to $\sim$30$\%$. Both the $o$-H$_2$O 1$_{01}$ and $p$-H$_2$O 0$_{00}$
                 ground-state levels are at $E_{\rm low}/k$=0\,K.}
\centering
\begin{tabular}{ccccc}
\hline\hline
\multicolumn{1}{c}{Species} & 
\multicolumn{1}{c}{Transition} &
\multicolumn{1}{c}{E$_{up}$} &
\multicolumn{1}{c}{$\lambda$} &
\multicolumn{1}{c}{Line Flux}  \\
         &   & (K) &  ($\mu$m) & (W\,m$^{-2}$) \\
\hline\hline
%  	    &	 			& 	  &    & \\
HCO$^+$     &	5 - 4   		& 64.2	  & 672.326   & 2.24E-16\\
CO  	    &	4 - 3 			& 55.3 	  & 650.252   & 4.11E-16\\
C           &	$^3P_1$-$^3P_0$ 	& 23.6    & 609.133   & 8.69E-17\\
HCO$^+$     &	6 - 5   		& 89.9	  & 560.294   & 3.16E-17\\
$^{13}$CO   &	5 - 4 			& 79.3    & 544.161   & 6.33E-17\\
H$_2$O      &	1$_{10}$-1$_{01}$ 	& 26.7    & 538.289   & 4.96E-17\\
CO          &	5 - 4 			& 83.0	  & 520.231   &	5.13E-16\\
HCO$^+$     &	7 - 6   		& 119.8	  & 480.275   & 6.20E-17\\
$^{13}$CO   &	6 - 5 			& 111.1	  & 453.498   &	1.11E-16\\
CO          &	6 - 5 			& 116.2	  & 433.556   &	6.85E-16\\
HCN         &	8 - 7   		& 153.1	  & 422.912   & 3.28E-17\\
HCO$^+$     &	8 - 7   		& 154.1	  & 420.263   & 5.51E-17\\
H$_2$O      &	2$_{11}$-2$_{02}$	& 136.9	  & 398.643   & 1.67E-16\\
$^{13}$CO   &	7 - 6 			& 148.1	  & 388.743   &	1.42E-16\\
HCO$^+$     &	9 - 8   		& 192.6	  & 373.591   & 4.10E-17\\
CO          &	7 - 6  			& 154.9	  & 371.650   &	1.32E-15\\
C           &	$^3P_2$-$^3P_1$ 	&  62.5	  & 370.414   & 1.42E-16\\
$^{13}$CO   &	8 - 7 			& 190.4	  & 340.181   &	4.16E-17\\
HCO$^+$     &  10 - 9   		& 235.4	  & 336.255   & 3.37E-17\\
%H$_2$O      &	4$_{22}$-2$_{31}$	& 454.4	  & 327.223   &	7.58E-16\\
CO          &	8 - 7 			& 199.1	  & 325.225   & 1.68E-15\\
HCN         &  11 - 10   		& 280.7	  & 307.641   & 5.55E-17\\
HCO$^+$     &  11 - 10   		& 282.4	  & 305.712   & 8.91E-17\\
H$_2$O      &	2$_{02}$-1$_{11}$	& 100.8	  & 303.459   &	4.40E-16\\\hline
$^{13}$CO   &	9 - 8 			& 237.9	  & 302.415   &	1.59E-16\\
CO          &	9 - 8 			& 248.9	  & 289.120   & 1.54E-15\\
HCO$^+$     &  12 - 11   		& 333.8	  &  280.257  & 8.18E-17\\
H$_2$O      &	3$_{12}$-3$_{03}$	& 215.2	  & 273.193   &	3.23E-16\\
$^{13}$CO   &	10 - 9 			& 290.8   & 272.205   &	5.62E-17\\
H$_2$O      &	1$_{11}$-0$_{00}$	& 53.4	  & 269.273   &	4.41E-16\\
CO          &	10 - 9 			& 304.2	  & 260.240   &	2.02E-15\\
H$_2$O      &	3$_{21}$-3$_{12}$	& 271.0	  & 257.790   &	2.48E-16\\
$^{13}$CO   &	11 - 10	 		& 348.9	  & 247.490   & 1.51E-16\\
H$_2$O      &	2$_{20}$-2$_{11}$	& 195.9	  & 243.972   & 2.92E-17\\
CO          &	11 - 10			& 365.0	  & 236.613   &	2.17E-15\\
$^{13}$CO   &	12 - 11			& 412.4	  & 226.898   &	7.67E-17\\
CO          &	12 - 11			& 431.3	  & 216.927   &	2.45E-15\\
$^{13}$CO   &	13 - 12		        & 481.0	  & 209.476   &	1.51E-16\\
CO          &	13 - 12			& 503.6	  & 200.272   &	2.60E-15\\\hline\hline
$^{13}$CO   &	14 - 13			& 555.0	  & 194.546   &	1.32E-16\\
H$_2$O      &	4$_{13}$-4$_{04}$	& 396.4	  & 187.110   &	5.74E-17\\
CO          &	14 - 13			& 581.0	  & 185.999   &	1.55E-15\\
$^{13}$CO   &	15 -14			& 634.2	  & 181.608   &	6.57E-17\\
H$_2$O      &	2$_{21}$-2$_{12}$	& 159.9	  & 180.488   &	5.25E-16\\
H$_2$O      &	2$_{12}$-1$_{01}$	&  80.1	  & 179.527   &	9.72E-16\\
H$_2$O      &	3$_{03}$-2$_{12}$	& 162.5	  & 174.626   &	1.21E-15\\
CO          &	15 - 14			& 663.9	  & 173.631   &	2.20E-15\\
$^{13}$CO   &	16 - 15			& 718.7	  & 170.290   &	3.63E-17\\
OH          &	$^{2}\Pi_{1/2}$ $J$=3/2$^-$-1/2$^+$ & 269.8   & 163.396  & 1.92E-16\\
OH          &	$^{2}\Pi_{1/2}$ $J$=3/2$^+$-1/2$^-$ & 270.2   & 163.015  & 2.76E-16\\
CO          &	16 - 15			& 752.4	  & 162.812   &	2.01E-15\\
U           &	 			& 	  & 161.808$^a$   &	3.49E-17\\
C$^+$       &	$^2P_{3/2}$-$^2P_{1/2}$ & 91.2	  & 157.741   &	2.92E-16\\
H$_2$O      &	3$_{22}$-3$_{13}$	& 296.8	  & 156.194   &	3.16E-16\\
CO          &	17 - 16			& 846.3	  & 153.267   &	2.15E-15\\
O           &	$^3P_0$-$^3P_1$		& 326.6	  & 145.525   &	8.38E-16\\
CO          &	18 - 17			& 945.8	  & 144.784   &	2.04E-15\\
H$_2$O      &	3$_{13}$-2$_{02}$	& 204.7	  & 138.527   &	9.49E-16\\\hline\hline
\end{tabular}
\tablefoottext{a}{Observed wavelength (otherwise rest wavelengths are tabulated)}
\end{table}

 \begin{table}[h]
      \caption[]{Continuation of Table~1.}
\centering
\begin{tabular}{ccccc}
\hline\hline
\multicolumn{1}{c}{Species} & 
\multicolumn{1}{c}{Transition} &
\multicolumn{1}{c}{E$_{up}$} &
\multicolumn{1}{c}{$\lambda$} &
\multicolumn{1}{c}{Line Flux}  \\
         &   & (K) &  ($\mu$m) & (W\,m$^{-2}$) \\
\hline\hline
CO          &	19 - 18			& 1050.7  & 137.196   &	1.66E-15\\
H$_2$O      &	5$_{14}$-5$_{05}$	& 540.5	  & 134.935   &	1.67E-16\\
U           &	 			& 	  & 132.982$^a$   &	1.13E-16\\
U           &	 			& 	  & 132.672$^a$   &	1.76E-16\\
H$_2$O      &	4$_{23}$-4$_{14}$	& 397.9	  & 132.407   &	3.95E-16\\
CO          &	20 - 19			& 1161.2  & 130.369   &	1.62E-15\\
H$_2$O      &	4$_{04}$-3$_{13}$	& 319.5	  & 125.353   &	6.79E-16\\
CO          &	21 - 20			& 1277.1  & 124.193   &	2.01E-15\\
OH          &	$^{2}\Pi_{3/2}$ $J$=5/2$^+$-3/2$^-$ & 120.5	  & 119.441   &	8.18E-16\\
OH          & 	$^{2}\Pi_{3/2}$ $J$=5/2$^-$-3/2$^+$ & 120.8	  & 119.234   &	8.75E-16\\
CO          &	22 - 21				    & 1398.6      & 118.581   &	1.29E-15\\
H$_2$O      &	4$_{14}$-3$_{03}$		    & 289.3	  & 113.537   &	7.39E-17$^{\dagger}$\\
CO          &	23 - 22				    & 1525.5      & 113.458   &	blended$^{\dagger}$\\
CO          &	24 - 23				    & 1657.9      & 108.763   &	1.08E-15\\
H$_2$O      &	2$_{21}$-1$_{10}$		    & 159.9	  & 108.073   &	1.35E-16\\
CO          &	27 - 26	                	    & 2087.9      & 96.773    &	9.06E-16\\
U           &	 	                	    &       	  & 96.633$^a$    & 6.14E-17\\
OH          &	$^{2}\Pi_{1/2}$-$^{2}\Pi_{3/2}$ $J$=3/2$^-$-5/2$^+$ & 269.8       & 96.363    & 2.35E-16\\
OH          &	$^{2}\Pi_{1/2}$-$^{2}\Pi_{3/2}$ $J$=3/2$^+$-5/2$^-$ & 270.2       & 96.271    & 2.35E-16\\
H$_2$O      &	5$_{15}$-4$_{04}$		    & 470.0	  & 95.626    &	5.16E-16\\
H$_2$O      &	4$_{41}$-4$_{32}$ 	            & 668.1       & 94.703    & 1.76E-16\\
CO          &	28 - 27				    & 2242.2  	  & 93.349    &	7.55E-16\\
U           &                                       &             & 90.574$^a$    & 4.19E-17 \\
U           &                                       &             & 90.434$^a$    & 4.64E-18 \\
CO          &	29 - 28				    & 2401.9      & 90.163    &	6.57E-16\\
H$_2$O      &	3$_{22}$-2$_{11}$	            & 296.8	  & 89.988    &	8.29E-16\\
CO          & 	30 - 29 			    & 2567.0  	  & 87.190    & 7.03E-16\\
U           &                                       &             & 86.944$^a$& 6.48E-17 \\
H$_2$O      &   7$_{16}$-7$_{07}$                   & 979.0       & 84.766    & 1.06E-16\\
OH          &   $^{2}\Pi_{3/2}$ $J$=7/2$^-$-5/2$^+$ & 290.5 	  & 84.597    &	9.29E-16\\
CO          &	31 - 30				    & 2737.6      & 84.411    &	1.25E-15$^{\dagger}$\\
OH          &   $^{2}\Pi_{3/2}$ $J$=7/2$^+$-5/2$^-$ & 291.2       & 84.420    &	blended$^{\dagger}$    \\
H$_2$O      &	6$_{06}$-5$_{15}$		    & 642.7	  & 83.283    &	1.69E-16\\
H$_2$O      &	6$_{16}$-5$_{05}$		    & 609.3	  & 82.031    &	5.55E-16\\
CO          &	32 - 31 			    & 2913.7      & 81.806    &	3.84E-16\\
U           &                                       &             & 81.013$^a$& 7.03E-17 \\
CO          &	33 - 32				    & 3095.1      & 79.360    &	2.49E-16\\
OH          &	$^{2}\Pi_{1/2}$-$^{2}\Pi_{3/2}$ $J$=1/2$^+$-3/2$^-$ & 181.7 & 79.179 &	blended$^{\dagger}$\\
OH          &	$^{2}\Pi_{1/2}$-$^{2}\Pi_{3/2}$ $J$=1/2$^-$-3/2$^+$ & 181.9 & 79.116 &	9.97E-16$^{\dagger}$\\
H$_2$O      &	4$_{23}$-3$_{12}$		    & 397.9	  & 78.741    &	4.89E-17\\
CO          &	34 - 33				    & 3282.0      & 77.059    &	3.97E-16\\
H$_2$O      &	3$_{21}$-2$_{12}$ 		    &  271.0  	  & 75.380    &	7.23E-16\\
CO          &	35 - 34				    & 3474.3      & 74.890    &	3.53E-16\\
U           &                                       &             & 69.840$^a$    & 5.75E-17\\
CO          &	38 - 37				    & 4083.5      & 69.074    &	1.57E-16\\
CO          &	39 - 38				    & 4297.3      & 67.336    &	1.28E-16\\
H$_2$O      &	3$_{03}$-3$_{03}$		    & 376.4	  & 67.269    &	7.77E-17\\
H$_2$O      &	3$_{31}$-2$_{20}$		    & 410.4	  & 67.089    &	1.90E-16\\
H$_2$O      &	3$_{30}$-2$_{21}$		    & 376.4	  & 66.437    &	4.74E-16\\
H$_2$O      &	7$_{16}$-6$_{25}$		    & 979.0	  & 66.092    & 1.28E-16\\
CO          &	40 - 39				    & 4516.6 	  & 65.686    &	8.33E-17\\
OH          &	$^{2}\Pi_{3/2}$ $J$=9/2$^+$-7/2$^-$ & 511.0	  & 65.279    &	4.45E-16\\
H$_2$O      &	6$_{25}$-5$_{14}$ 		    & 761.3	  & 65.165    &	1.99E-16\\
OH          &	$^{2}\Pi_{3/2}$ $J$=9/2$^-$-7/2$^+$ & 512.1	  & 65.132    &	4.11E-16\\
CO          &	41 - 40				    & 4741.2      & 64.117    &	1.23E-16\\
H$_2$O      &   8$_{08}$-7$_{17}$                   & 1070.6      & 63.457    & 2.81E-17\\
H$_2$O      &   8$_{18}$-7$_{07}$                   & 1036.5      & 63.322    & 4.00E-17\\
O           &	$^3P_1$-$^3P_2$			    & 227.7	  & 63.183    &	8.34E-15\\
CO 	    &	42 - 41				    & 4971.1      & 62.624    &	6.82E-17\\
U           &                                       &             & 62.003$^a$ & 9.05E-17\\
H$_2$O      &	7$_{26}$-6$_{15}$       	    & 1021.8      & 59.986    &	1.40E-16\\
H$_2$O      &	4$_{32}$-3$_{21}$		    & 516.1	  & 58.698    &	2.59E-16\\
H$_2$O      &	4$_{22}$-3$_{13}$		    & 454.4	  & 57.636    &	2.64E-16\\
H$_2$O      &   4$_{31}$-3$_{22}$                   & 552.3       & 56.324    & 1.05E-16\\
U           &                                       &             & 55.481$^a$    & 8.01E-17\\\hline\hline
\end{tabular}
\tablefoottext{a}{Observed wavelength (otherwise rest wavelengths are tabulated)}
\end{table}

\clearpage

 \begin{table}[t]
      \caption[]{Mid-IR lines detected with \textit{Spitzer}/IRS towards the Class\,0 
                 protostar Serpens SMM1 and 
                 line fluxes (uncorrected for extinction) within a 7$''$$\times$7$''$ aperture.
                 The horizontal line shows the transition between the LH and SH modules.
                 Intensity errors up to $\sim$25$\%$.}
\centering
\begin{tabular}{ccccc}
\hline\hline
\multicolumn{1}{c}{Species} & 
\multicolumn{1}{c}{Transition} &
\multicolumn{1}{c}{E$_{up}$} &
\multicolumn{1}{c}{$\lambda$} &
\multicolumn{1}{c}{Line Flux}  \\
            &                 & (K)       &  ($\mu$m) & (W\,m$^{-2}$) \\
\hline\hline
%  	    &	 			& 	  &    & \\
Fe$^+$      &	$^6D_{5/2}$-$^6D_{7/2}$ &   961	  & 35.349   & 2.45E-17\\
Si$^+$      &	$^2P_{3/2}$-$^2P_{1/2}$ &   414   & 34.815   & 6.29E-17\\
$p$--H$_2$ &   0-0 $S$(0) $J$=2--0      &   509	  & 28.219   & $<$5.92E-18 \\
Fe$^+$      &	$^6D_{7/2}$-$^6D_{9/2}$ &   554	  & 25.988   & 8.12E-17\\
S        &	$^3P_1$-$^3P_2$         &   570	  & 25.249   & 3.24E-17\\
Fe$^+$      &	$^4F_{5/2}$-$^4F_{7/2}$ &  4087	  & 24.519   & 4.46E-17\\\hline
Fe$^+$      &	$^4F_{7/2}$-$^4F_{9/2}$ &  3499	  & 17.936   & 1.36E-17\\
$o$--H$_2$ & 0-0 $S$(1) $J$=3--1        &  1015	  & 17.035   & 5.96E-19\\
Ne$^+$      & $^2P_{1/2}$-$^2P_{3/2}$    &  1124	  &  12.814  & 7.06E-18\\
$p$--H$_2$ &0-0 $S$(2) $J$=4--2         &  1682	  &  12.279  & 2.42E-18 \\\hline\hline
\end{tabular}
\end{table}

\begin{figure}[t]
\hspace{3cm}
\begin{center}
\resizebox{\hsize}{!}{\includegraphics[angle=-0]{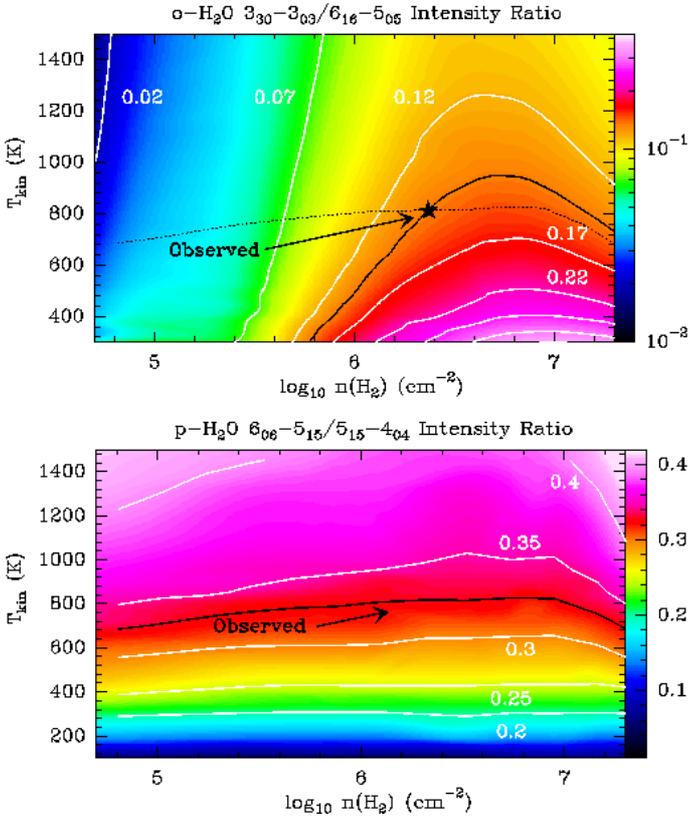}}
\caption{Grid of H$_2$O LVG models for different gas temperatures and densities.
\textit{Bottom}: Contour levels of the $p$-H$_2$O 6$_{06}$-5$_{15}$/5$_{15}$-4$_{04}$ (83.284/95.627\,$\mu$m)
line ratio for $N$($p$-H$_2$O)=10$^{16}$\,cm$^{-2}$. Note the gas temperature dependence of this ratio.
The black continuous curve shows the observed  intensity line ratio of 0.33.
\textit{Top}: Contour levels of the $o$-H$_2$O 3$_{30}$-3$_{03}$/6$_{15}$-3$_{05}$ (67.269/82.031\,$\mu$m)
ratio for  $N$($o$-H$_2$O)=3$\times$10$^{16}$\,cm$^{-2}$. Note the gas density dependence of this ratio.
The black continuous curve shows the observed  intensity line ratio of 0.14 while the black dotted curve
shows the observed $p$-H$_2$O 6$_{06}$-5$_{15}$/5$_{15}$-4$_{04}$ ratio.
The intersection of both curves is marked with a star.} 
\end{center}
\label{fig_h2ogrid}
\end{figure}

\begin{figure}[t]
\begin{center}
\resizebox{\hsize}{!}{\includegraphics[angle=-90]{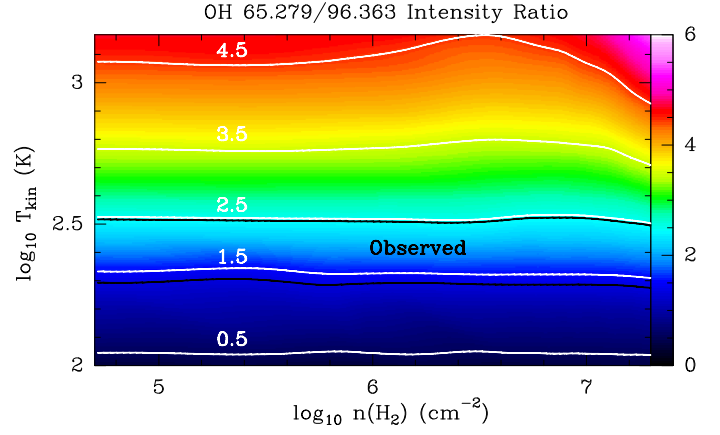}}
\caption{Grid of OH LVG models for different gas temperatures and densities.
Contour levels of the OH $^{2}\Pi_{3/2}$ $J$=9/2$^+$-7/2$^-$/$^{2}\Pi_{1/2}$-$^{2}\Pi_{3/2}$ $J$=3/2$^-$-5/2$^+$ (65.279/96.363\,$\mu$m)
line ratio for $N$(OH)=10$^{16}$\,cm$^{-2}$. Note the gas temperature dependence of this ratio.
The black continuous curves enclose the observed  intensity line ratio with a $\sim$15\% of relative
flux calibration error.} 
\end{center}
\label{fig_ohgrid}
\end{figure}

\clearpage

\begin{figure*}[t]
\begin{center} 
\includegraphics[width=11cm, angle=-90]{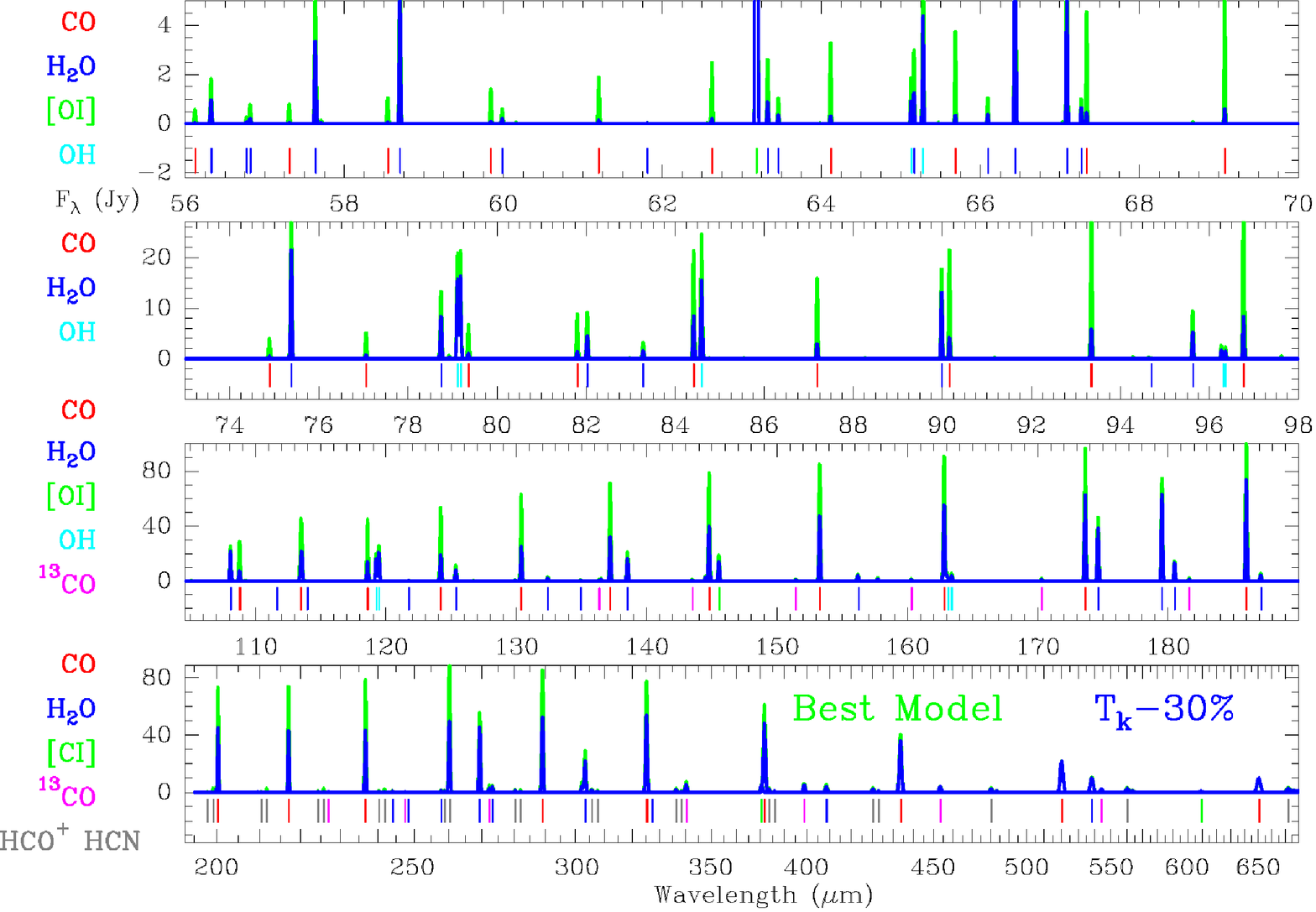}
\caption{Best model discussed in the text (green curves) and a model where T$_{\rm k}$ 
 is decreased by 30\% in all components (blue curves).} 
\end{center}
\label{fig_modeltemp}
\end{figure*}

\begin{figure*}[b]
\begin{center} 
\includegraphics[width=11cm, angle=-90]{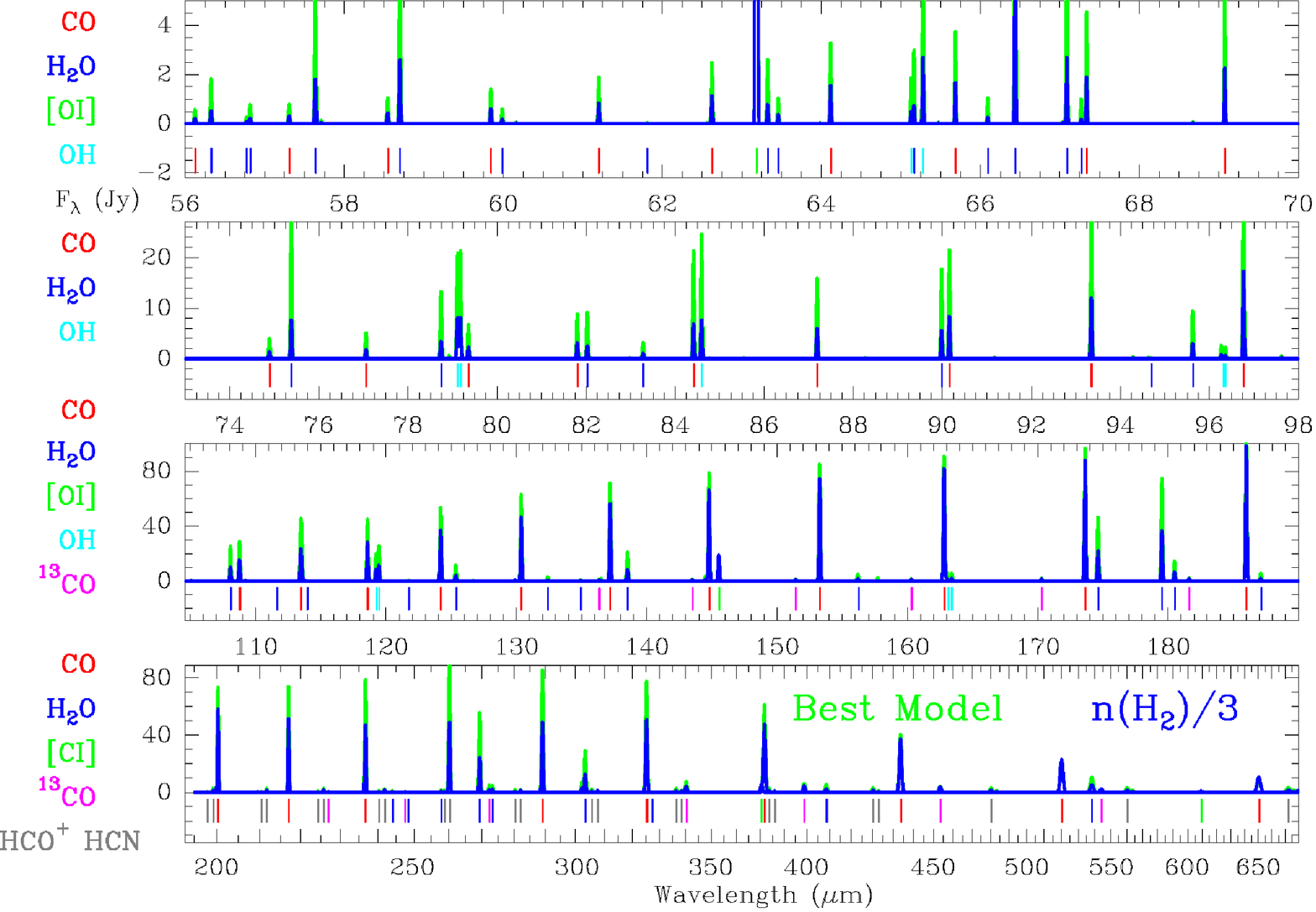}
\caption{Best model discussed in the text (green curves) and a model where  
 $n$(H$_2$) is decreased by a factor 3 in all components (blue curves).} 
\end{center}
\label{fig_modeldens}
\end{figure*}

\end{appendix}
\end{document}